\documentclass[referee]{raa}
\usepackage{graphicx,times}
\usepackage{natbib}

\usepackage{amssymb,amsmath}
\bibpunct{(}{)}{;}{a}{}{,}

\usepackage[pagebackref=true]{hyperref}

\begin{document}

   \title{Research Progress on Solar Small-Scale Dynamo}

 \volnopage{ {\bf 20XX} Vol.\ {\bf X} No. {\bf XX}000--000}
   \setcounter{page}{1}

   \author{Wen-Jie Jiang  
   \inst{1,2}, Lei Ni\inst{1,2,4*}, Chun-Lan Jin\inst{2,3*}, Zhi Xu\inst{1,2,4*}, Mei Zhang\inst{2,3*}
   }

   \institute{ Yunnan Observatories, Chinese Academy of Sciences, Kunming, Yunnan 650216, P.R.China; {\it leini@ynao.ac.cn}, {\it xuzhi@ynao.ac.cn}\\
\and
University of Chinese Academy of Sciences, Beijing 100049, P.R.China\\
\and
State Key Laboratory of Solar Activity and Space Weather, National Astronomical Observatories, Chinese Academy of Sciences, Beijing 100101, P.R.China; {\it cljin@nao.cas.cn}, {\it zhangmei@bao.ac.cn}\\
\and
Yunnan Key Laboratory of Solar Physics and Space Science, Kunming 650216, P.R.China\\
\vs \no
   {\small Received 20XX Month Day; accepted 20XX Month Day}
}

\abstract{The small-scale solar dynamo theory, as the core mechanism explaining the origin of the persistent, disperse weak magnetic field in the quiet Sun regions, has made significant progress over the past three decades in the fields of observation, theory, and simulation. Breakthrough observations from high-resolution space-based and ground-based telescopes (e.g.Hinode, SDO/HMI, SUNRISE, SST, GREGOR, GST, DKIST) have revealed that the quiet Sun is ubiquitously populated by highly dynamic, mixed-polarity and possibly predominantly horizontal magnetic structures with complex topology. These observations confirm that their total magnetic flux is substantial, with a high and widely distributed magnetic flux emergence rate, strongly suggesting a local dynamo effect{-}independent of the solar cycle and driven by intense turbulence and convection. Theoretical studies indicate that even in the challenging low magnetic Prandtl number ( $P_m\ll$ 1) environment of the solar photosphere and convection zone, turbulent motions can self-excitedly convert kinetic energy into magnetic energy through the stretching, folding, and twisting of magnetic field lines. Magnetohydrodynamic (MHD) numerical simulations have successfully reproduced observed features, demonstrating that a pure small-scale dynamo can operate efficiently and sustain magnetic fields even in an open, stratified solar environment incorporating realistic physical processes (e.g.radiative transfer, partial ionization, etc.). Current research strongly suggests that small-scale magnetic fields constitute the majority of the magnetic energy in the quiet Sun and also influence coronal heating, solar wind acceleration, and radiation distribution. This article conducts a literature review centered on observations, theoretical models, and numerical simulations of the small-scale dynamo, organizing and discussing the relevant research history and progress. Finally, it summarizes the content and provides an outlook on future research from multiple perspectives.
\keywords{Solar small-scale dynamo---Quiet Sun---Magnetohydrodynamics (MHD)}
}

   \authorrunning{W.-J. Jiang et al. }            
   \titlerunning{Research Progress on Solar Small-Scale Dynamo}  
   \maketitle

%
\section{Introduction}           
\label{sect:intro}

The Sun, as a typical magnetically active star, possesses a ubiquitous magnetic field permeating its surface and atmosphere, which serves as the fundamental engine driving the dynamic changes in the heliospheric space environment. From structuring the corona and heating the million-degree plasma \citep{1998Natur.394..152S}to triggering solar flares and coronal mass ejections that impact the Earth's space weather \citep{2019ApJ...871...16J} and to modulating the total solar irradiance (TSI) and spectral solar irradiance (SSI) which potentially influence the Earth's climate \citep{2014ApJ...789..132R}, solar magnetic activity profoundly affects interplanetary space and human society. Understanding the origin, evolution, and energy release mechanisms of the solar magnetic field is thus a central goal of solar physics. Traditional research on solar magnetic activity has predominantly focused on large-scale phenomena or processes, such as sunspots, active regions, and the reversal of the global dipole field. The origin of these large-scale magnetic fields is widely attributed to the global dynamo operating deep within the solar interior (primarily at the base of the convection zone and the tachocline). Its physical foundation lies in the solar differential rotation (the $\Omega$-effect) stretching the poloidal field, transforming it into a toroidal field, followed by convective motions (the $\alpha$-effect) converting this toroidal field back into a new poloidal field, thereby forming a self-sustaining oscillatory cycle \citep{2013IAUS..294...37C,2011ApJ...733...90D,2022ApJ...933..199H,2014SSRv..186..491J,1955ApJ...121..491P}. 

\begin{figure}[!ht]
   \centering
   \includegraphics[width=0.9\textwidth]{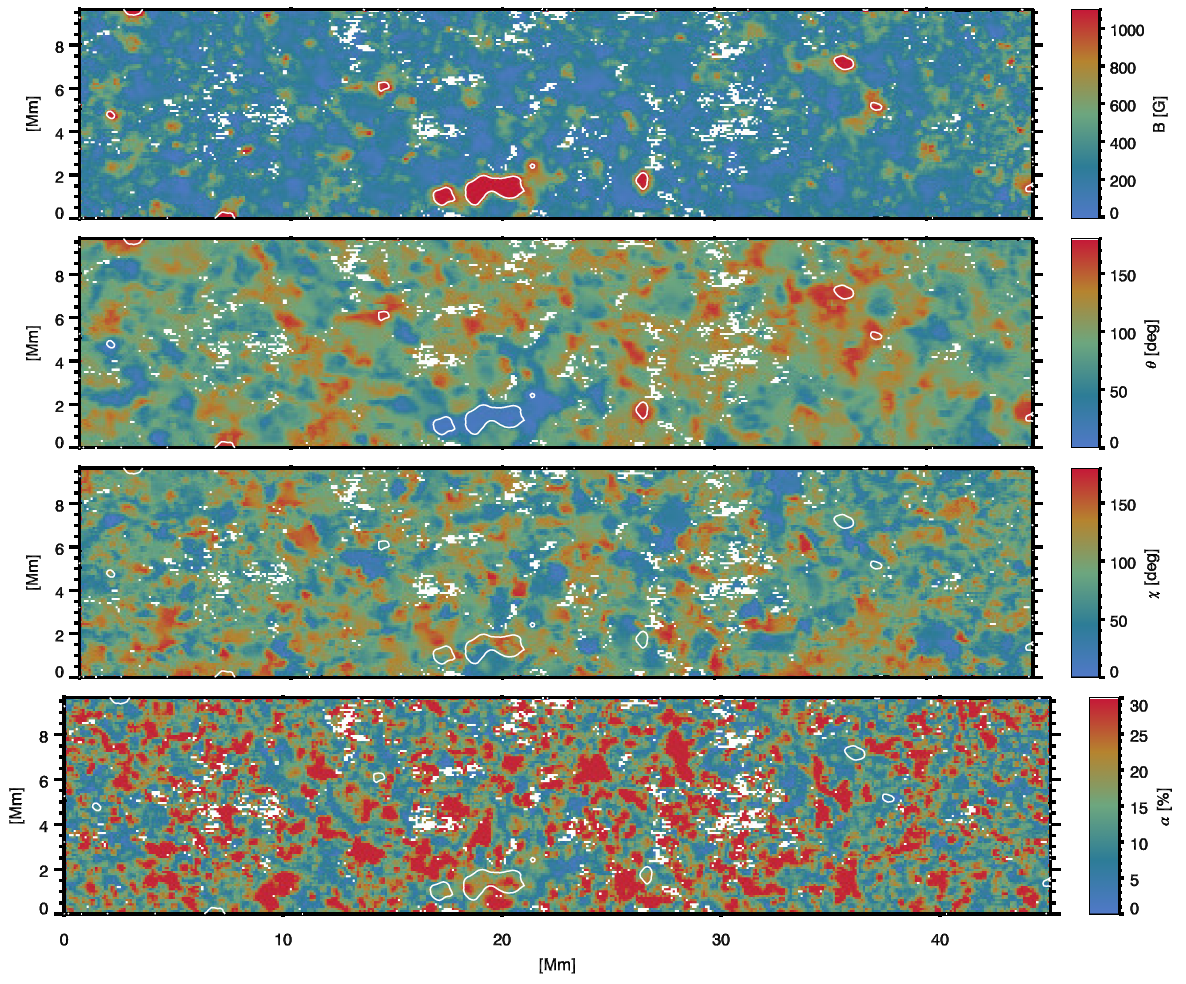}
   \caption{ Distribution of the magnetic field in the quiet Sun derived from denoised observational data obtained by the GREGOR Infrared Spectrograph (GRIS) on the GREGOR telescope, reproduced from \citet{2016A&A...596A...5M}.. From top to bottom, the panels show the strength of the vector magnetic field, its inclination angle, its azimuth angle, and the magnetic filling factor, respectively. The white contours indicate regions where the magnetic field strength corresponds to the kilogauss line. White pixels represent areas where the signal value is below 4$\sigma$n and were therefore not inverted.\newline \small Credit: M. J. Martínez González et al. \textit{A\&A},Vol. 596, A5, 2016, Fig. 2. DOI: \href{http://www.aanda.org/10.1051/0004-6361/201628449}{10.1051/0004-6361/201628449}. Reproduced with permission © ESO.}
   \label{fig1}
   \end{figure}

However, the rapid progress on the high-resolution space and ground based observations, particularly since the 21st century with missions and facilities like the HSOS Solar Multi-Channel Telescope \citep{1999A&A...352..317Z, 2000SoPh..196..269Z},  the Hinode satellite \citep{2008ApJ...684.1469D,2008ApJ...672.1237L,2007ApJ...670L..61O}, the SUNRISE balloon-borne solar telescope \citep{2013SSRv..178..141M,2010ApJ...723L.127S}, the American Goode Solar Telescope (GST) at the Big Bear Lake \citep{2010AN....331..636C}, the Swedish 1-m Solar Telescope (SST) \citep{2003SPIE.4853..341S},  the German 1.5-meter GREGOR telescope \citep{2015LRCA....1....2S}, and the recently operational Daniel K. Inouye Solar Telescope (DKIST) in the USA \citep{2017nsf....1718947M}, has revealed a more ubiquitous and dynamic magnetic world in the Quiet Sun. In the seemingly tranquil non-active regions, far from sunspots, the Quiet Sun is permeated by complex, mixed-polarity, small-scale ($\sim1$\,Mm) magnetic structures with lifetimes ranging from minutes to tens of minutes. The magnetic flux of individual small-scale structures is typically small ($< 10^{19}$ Mx, far less than the typical sunspot flux of $10^{21}-10^{22}$ Mx) with field strengths ranging from tens of Gauss to several kilogauss \citep{2013A&A...555A..33B,1962AEEP...16..431L,2003ApJ...597L.177S,2012A&A...541A..17S}. Although individually weak, these small-scale magnetic elements are vast in number, widely distributed spatially, and exhibit extremely rapid renewal rates \citep{2003ApJ...584.1107H}. During solar minimum, the quiet Sun contributes over 90\% of the Sun's total magnetic flux \citep{2011ApJ...731...37J}. Even at solar maximum, the flux contribution from quiet Sun is also comparable to ARs \citep{2019RAA....19...69J,2026ApJS..283...33J}.Figure \ref{fig1} shows the distribution of the magnetic field in the quiet Sun based on denoised GRIS observational data \citep{2016A&A...596A...5M}. 

These seemingly "insignificant" small-scale magnetic fields are the key energy sources for heating the corona, particularly the quiet corona and coronal holes \citep{2007Sci...318.1572E,2019AGUFMSM13B..09L,2019Sci...366..890S,2014ApJ...797L..14T} and accelerating the solar wind \citep{2023Sci...381..867C,1998Natur.394..152S,2005Sci...308..519T} through energy dissipation mechanisms such as wave dissipation, magnetic reconnection, or nanoflares. They constitute the primary component of the photospheric magnetic field, directly influencing the distribution of the solar radiation field \citep{2011A&A...531A.112K,2012A&A...542A..96K,2020ApJ...894..140R} and serving as one important source for large-scale magnetic activities like solar flares.

The small-scale dynamo(SSD) is considered the key mechanism responsible for the origin of the persistent,  diffuse weak magnetic field in the quiet Sun \citep{1999ApJ...515L..39C,2024ApJ...961L..46C,1993A&A...274..543P,2023SSRv..219...36R,2007A&A...465L..43V}. Its core premise is that the intense, highly turbulent convective motions in the near-surface layer of the Sun can efficiently convert kinetic energy into magnetic energy through a turbulent dynamo or nonlinear MHD processes. This mechanism self-excitedly generates and sustains magnetic fields in local regions smaller than typical convective cells, without relying on large-scale shear (e.g.differential rotation) \citep{2005PhR...417....1B,2007NJPh....9..300S,2011arXiv1103.3138T}. The key physical processes involve turbulent stretching, folding, and twisting of magnetic field lines, overcoming magnetic diffusion to achieve exponential field growth. The high electrical conductivity (low magnetic diffusivity) in the solar photosphere results in a large magnetic Reynolds number ($R_m$) theoretically making turbulence readily capable of driving a dynamo. However, studying the small-scale dynamo within theoretical models that incorporate realistic solar photospheric conditions is exceptionally challenging. This is due to two primary factors: 1. They need to account for realistic physical effects such as gravitational stratification, strong radiative transfer, and partially ionized plasma effects \citep{2009ApJ...691..640R,2007A&A...465L..43V}. 2. The very low magnetic Prandtl number ($P_m$ = kinematic viscosity / magnetic diffusivity, $P_m \ll 1$) in the solar photosphere. This implies that the dissipation scale for fluid motions is much smaller than the magnetic diffusion scale. Consequently, both theoretical analysis and numerical simulations face tremendous difficulties in probing down to these exceedingly small scales to fully unravel the initiation of magnetic field stretching and amplification by turbulence that leads to the dynamo
effect \citep{2011ApJ...741...92B,2007PhRvL..98t8501I,2005ApJ...625L.115S}.

\begin{figure}[!ht]
   \centering
   \includegraphics[width=0.9\textwidth]{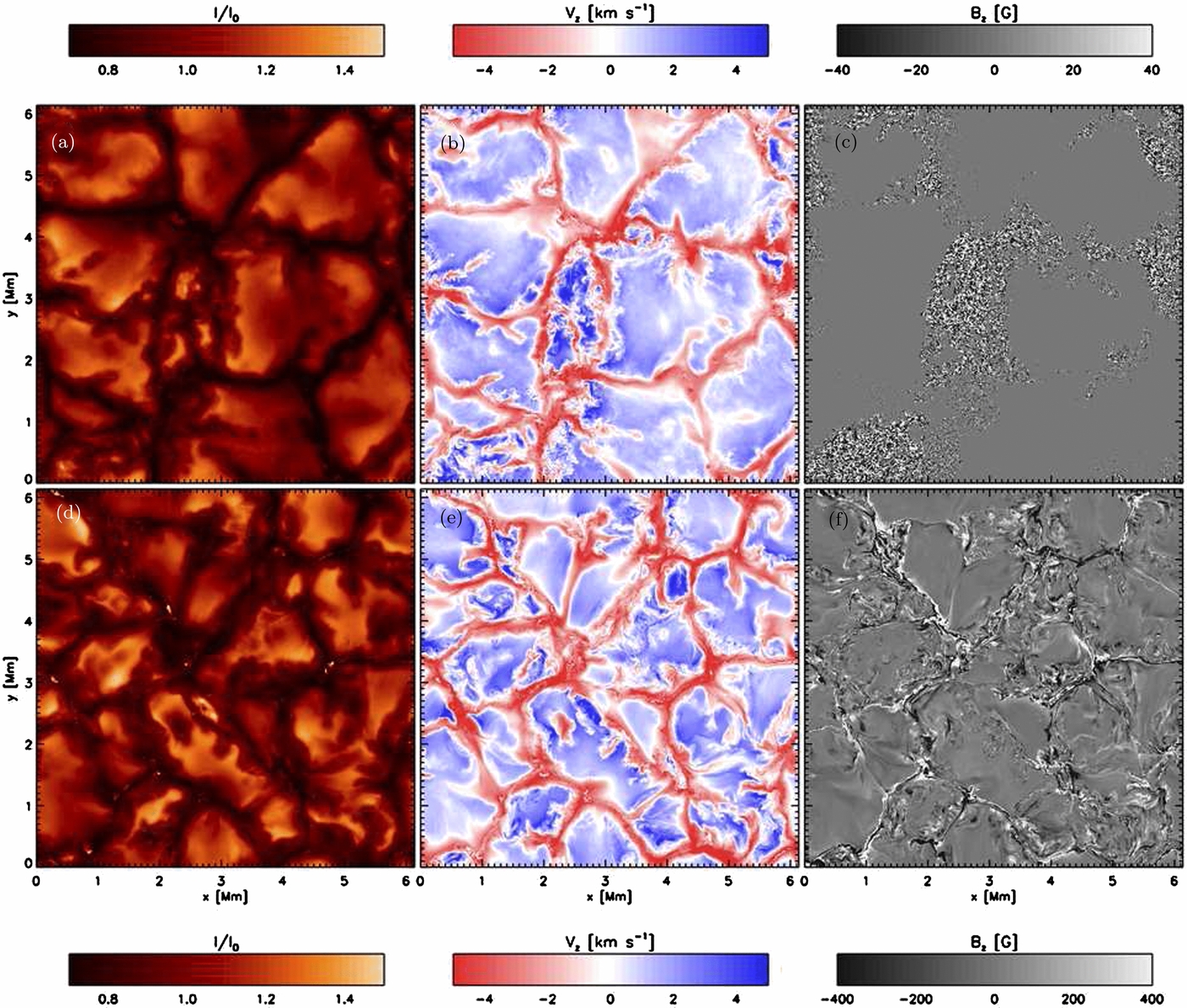}
   \caption{Numerical simulation results of small-scale dynamo from \citet{2014ApJ...789..132R}. Panels (a)–(c) show the kinematic growth phase, and panels (d)–(f) show the saturated phase. The left two panels display the intensity of the radiative heat flux. The middle two panels show the line-of-sight velocity at an optical depth of $\tau=1$. The right two panels show the distribution of the magnetic field component perpendicular to the solar surface at an optical depth of $\tau=1$. \newline \small Source: M. Rempel, ``Numerical Simulations of Quiet Sun Magnetism: On the Contribution from a Small-Scale Dynamo,'' \textit{The Astrophysical Journal}, Vol. 789, No. 2, Article 132, pp. 1–22, 2014 (Figure 1). DOI: \href{https://doi.org/10.1088/0004-637X/789/2/132}{10.1088/0004-637X/789/2/132}. © AAS. Reproduced with permission.}
   \label{fig2}
\end{figure}

Advances in numerical simulation techniques have provided a powerful tool for testing the small-scale dynamo theory and elucidating its physical mechanisms. Progress has been significant, evolving from early idealized simulations of incompressible turbulence \citep{2004PhRvL..92n4503P,2004Ap&SS.292..141S} to high-resolution, three-dimensional, local box simulations incorporating realistic solar physical conditions and processes (e.g.radiative transfer, partial ionization, realistic equation of state)\citep{2015ApJ...809...84K,2010ApJ...714.1606P,2014ApJ...789..132R,2007A&A...465L..43V}. These simulations have successfully reproduced many observed characteristics of small-scale magnetic fields \citep{2023SSRv..219...36R} such as mixed polarity, horizontal field dominance and the formation of magnetic bright points (MBPs). They confirm that under the solar photospheric conditions pure turbulent convection can effectively excite and sustain small-scale magnetic fields even in the absence of a large-scale seed field. Figure \ref{fig2} shows slices of various physical quantities from a small-scale dynamo simulation \citep{2014ApJ...789..132R}.

Further numerical simulations have probed the dynamical processes of flux emergence, such as the triggering process of magnetic buoyancy instability, the interaction between small-scale magnetic fields and convective structures, and the influence of small-scale magnetism on energy transport \citep{2014LRSP...11....3C,2011ApJ...735..126T}. Furthermore, the existence of the small-scale dynamo has also been identified in high-resolution global and semi-global large-scale simulations \citep{2014ApJ...786...24H,2016Sci...351.1427H,2017A&A...599A...4K}. However, numerous debates and unresolved questions persist. These concern on how the small-scale dynamo couples with the global dynamo, the contribution of the small-scale field to the solar cycle, and the precise quantification of the dynamo efficiency under the extremely low $P_m$ regime \citep{2005PhR...417....1B,2014ApJ...785...49P,2023SSRv..219...36R}.

Based on extensive literature research, this article will first focus on the physical properties of the quiet Sun magnetic field as revealed by high-resolution telescopes. It will then delve into the fundamental theoretical framework and the pivotal contributions of numerical simulations in elucidating the operating mechanisms of the small-scale dynamo. Finally, it will summarize the findings and outline future research directions. These prospects include leveraging next-generation observational facilities (e.g.DKIST, the under-construction 2.5-meter WeHoST, and the Chinese Giant Solar Telescope, CGST) combined with advanced numerical simulations. The article will also discuss the broader implications of solar small-scale dynamo research for understanding dynamo processes in other stars across the universe.


\section{Observational Studies of Solar Small-Scale Magnetic Fields}

The quiet Sun refers to regions of the solar surface without prominent activity phenomena such as sunspots and flares. Despite being termed "quiet," these regions are not truely magnetic vacuums but are instead filled with complex and dynamic small-scale magnetic structures. Although these magnetic fields are normally weak, they are crucial for understanding the energy balance of the solar atmosphere, coronal heating mechanisms, and the solar dynamo process. In recent years, breakthroughs in the observational study of solar small-scale magnetic fields have been achieved, thanks to developments in high-resolution space-based and ground-based telescopes (e.g.Hinode, SDO/HMI, SUNRISE, GREGOR, SST, GST, DKIST). This section will systematically outline the basic characteristics of the quiet Sun magnetic field, magnetic diagnostic techniques, and energy transport from an observational perspective.

\subsection{Characteristics of the Quiet Sun Magnetic Field}

\begin{figure}[!ht]
   \centering
   \includegraphics[width=0.9\textwidth,trim=100 50 100 0,clip]{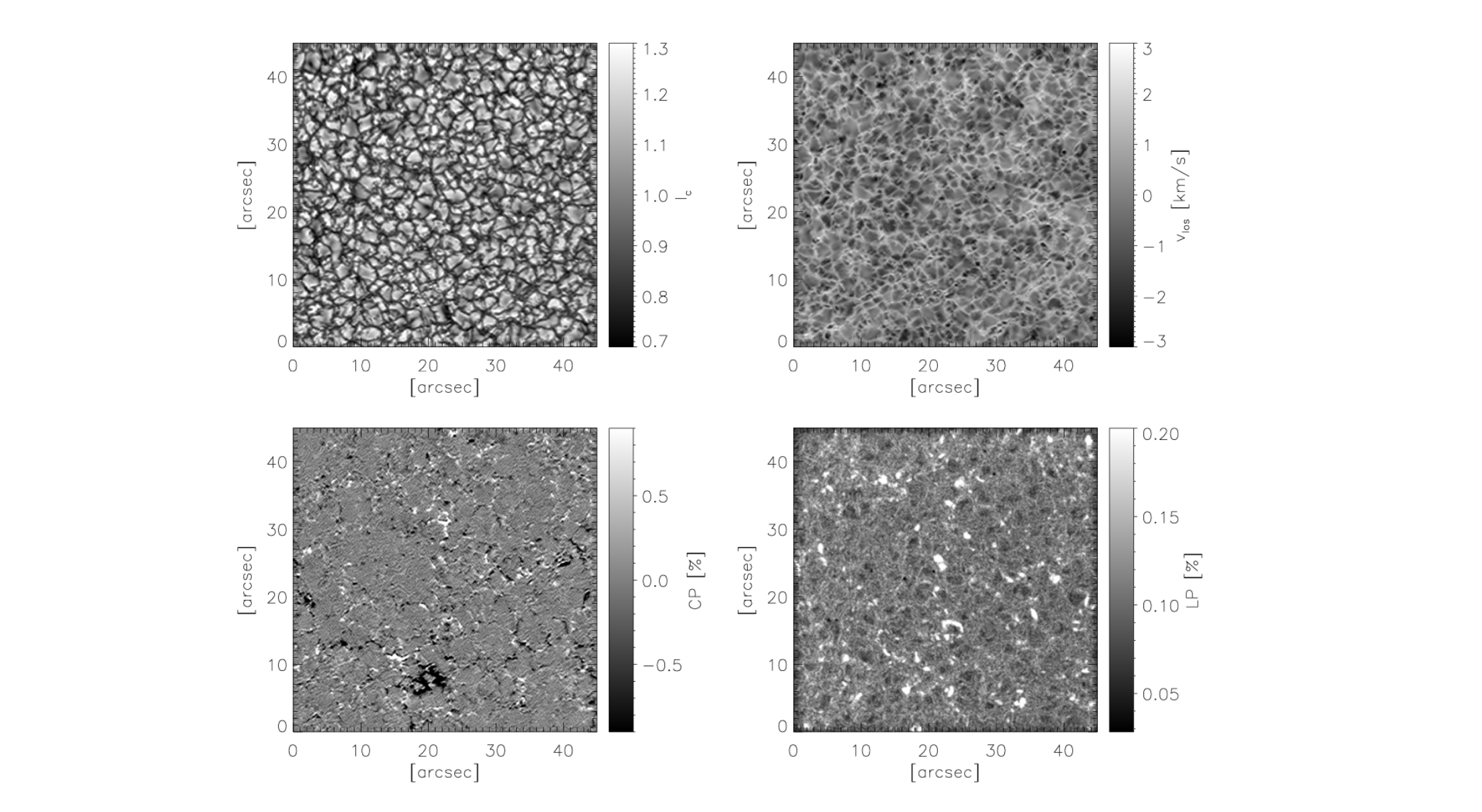}
   \caption{Quiet Sun region observed by SUNRISE/IMaX instrument, reproduced from \citet{2010ApJ...723L.127S}. The top-left panel displays the continuum intensity distribution at 5250.4 {\AA}. The top-right panel shows the line-of-sight velocity. The bottom-left panel presents the line-averaged Stokes V parameter map (related to the longitudinal magnetic field) derived from the Fe I 5250.2 {\AA} spectral line. The bottom-right panel reflects the net linear polarization (related to the transverse magnetic field). All images except for the linear polarization map are based on reconstructed data. \newline \small Source: S. K. Solanki et al., ``SUNRISE: Instrument, Mission, Data, and First Results,'' \textit{The Astrophysical Journal}, Vol. 723, No. 2, pp. L127–L133, 2010 (Figure 2). DOI: \href{https://doi.org/10.1088/2041-8205/723/2/L127}{10.1088/2041-8205/723/2/L127}. © AAS. Reproduced with permission.}
\label{fig:solanki_quiet_sun}
   \label{fig3}
   \end{figure}

The network-like structure, is the most important distribution characteristic of magnetic fields in the quiet regions of the Sun, with the later been classified into network magnetic fields and internetwork magnetic fields. Convective and turbulent motions in the interior of the sun are the primary causes of such magnetic structures. The network and internetwork magnetic fields exhibit significant differences in magnetic flux, spatial distribution, field orientation, and dynamics. Figure \ref{fig3} shows the quiet Sun as observed by SUNRISE/IMaX \citep{2010ApJ...723L.127S}.

The network magnetic fields are concentrated at the boundaries of supergranules and typically manifest as strong magnetic structures with high magnetic flux (around $10^{17} -10^{20}$ Mx), predominantly in the form of vertical components \citep{2010ApJ...723L.127S,1973SoPh...32...41S}. The Stokes V circular polarization signal in the bottom-left panel of Figure \ref{fig3} clearly reveals the network-like structures, with the largest, strong Stokes V signal patch located near the bottom center of this panel, indicating a prominent network magnetic element. The network field consists of magnetic flux tubes with field strengths of $1-2$\,kG and diameters of approximately $100-300$\,km \citep{2012ApJ...758L..40M}. Their lifetimes range from several hours to tens of hours \citep{2010ApJ...723L.164L,2014ApJ...796...79U}. These structures are transported from the supergranule interiors to the boundaries by flux transport processesand and dominate the magnetic flux in the quiet region \citep{2014ApJ...797...49G,2012ApJ...752..149I}, contributing $80-90\%$ of the magnetic flux in the quiet Sun region \citep{1995SoPh..160..277W,2014ApJ...797...49G}. Network regions are highly dynamic and evolve on supergranular timescales, likely involving sub-arcsecond small-scale processes such as fragmentation, emergence, cancellation, and coalescence \citep{2008ApJ...674..520L,2011SoPh..269...13T,2015ApJ...814..134I}. Observations have shown the coexistence of weak fields and kilogauss-level strong fields within small-scale network structures \citep{2004ApJ...616..587S}, where strong fields reside in small magnetic flux tubes and weak fields fill the space between them.

The internetwork magnetic field is distributed within the supergranular cells, generally with the magnetic fluxe below $10^{17}$\,Mx \citep{1995SoPh..160..277W, 1996ApJ...460.1019L,2007ApJ...670L..61O,2013SoPh..283..273Z}. The internetwork field is dominated by weak fields ($<300$\,G) and horizontal components. It appears highly intermittent \citep{2014A&A...572A..98A}, with magnetic flux being primarily carried by sub-arcsecond ephemeral magnetic loops \citep{2008A&A...479..229M,2007ApJ...666L.137C,2012ApJ...751....2O}. It is characterized by highly fragmented, small-scale magnetic loops with random orientations and a significant horizontal component \citep{2010ApJ...714L..94M,2013A&A...555A.132S}. These loops form rapidly through local magnetic emergence \citep{2007ApJ...666L.137C,2009ApJ...700.1391M} and have short lifetimes of only minutes \citep{2012ApJ...751....2O}. Internetwork regions cover approximately 90\% of the quiet Sun area \citep{2011A&A...527A..29B} and contribute about $10-20\%$ of the total quiet Sun magnetic flux \citep{1995SoPh..160..277W,2014ApJ...797...49G}. The internetwork magnetic field is highly dynamic and evolves very rapidly, with a magnetic flux appearance rate about $1000-10000$ times higher than that in active regions \citep{2016ApJ...820...35G,1994SoPh..150....1S,2017ApJS..229...17S}. Most of the Stokes V signals in Figure 3 are in fact attributable to the internetwork magnetic field, which consists of numerous, highly localized mixed-polarity magnetic patches. Inferences from the Stokes V profiles indicate that the characteristic sizes of internetwork magnetic features are generally smaller than 1 arcsecond, with many approaching the instrumental spatial resolution of about 0.15 arcsecond, highlighting their relatively small spatial scale. These properties of the internetwork field strongly suggest that their generation mechanism is related to the local small-scale dynamo \citep{2013A&A...555A..33B,2015ApJ...806..174J,2015ApJ...807...70J,2011ApJ...737...52L}.

The network and internetwork magnetic fields jointly constitute the two-component model of the quiet region magnetic field: the network field maintains the large-scale ordered structure and serves as the source region for small-scale jets/spicules and the solar wind \citep{2024NatAs...8.1246H,2014ApJ...797L..14T}; whereas the internetwork field is typically diffuse and disordered, contributes to the hidden magnetic energy in the photosphere\citep{2004Natur.430..326T} and influences chromospheric heating through magnetic reconnection \citep{2014ApJ...794..140V} .

\begin{figure}[!ht]
   \centering
   \includegraphics[width=0.9\textwidth]{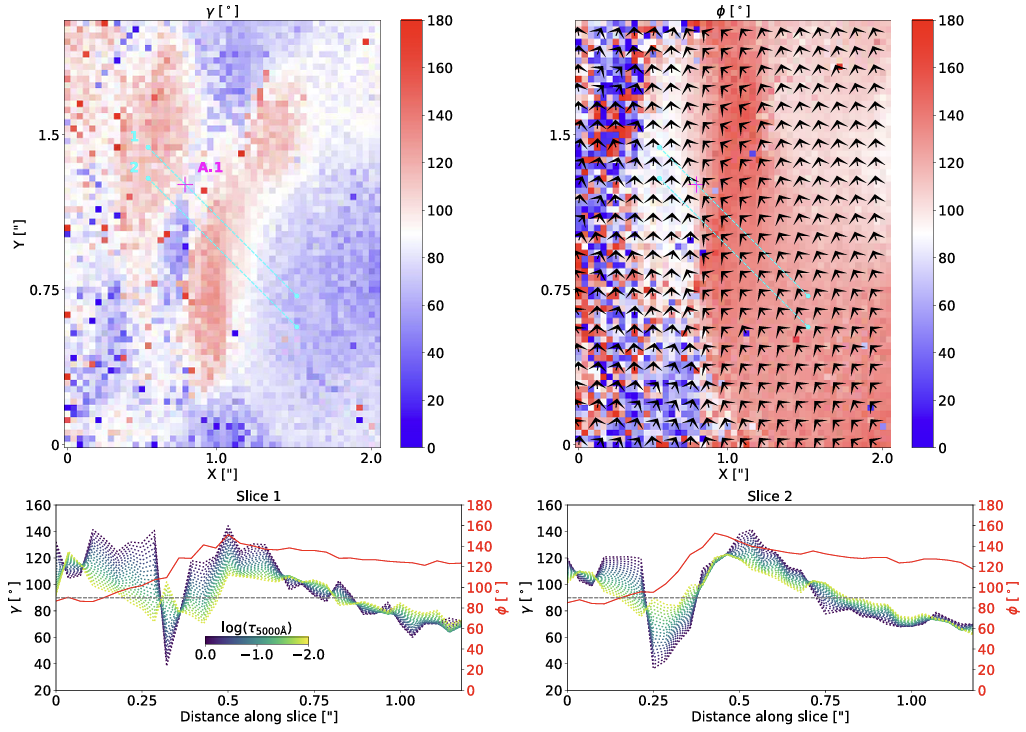}
   \caption{The top-left panel shows the spatial distribution of the magnetic inclination angle ($\gamma$) , reproduced from \citet{2023ApJ...955L..36C}. Within a scale of just 400 km, the $\gamma$ value crosses 90°three times, indicating three magnetic polarity reversals. The top-right panel displays the spatial distribution of the magnetic azimuth angle ($\phi$). The bottom two panels present a cross-sectional analysis, showing variations in magnetic field parameters along two paths (each path is about 882 km, the two light blue dashed lines in the top panels represent the two paths) tracing the serpentine structure. The distance from the first to the last polarity reversal is measured to be only about 400 km. \newline \small Source: R. J. Campbell et al., ``DKIST Unveils the Serpentine Topology of Quiet Sun Magnetism in the Photosphere,'' \textit{The Astrophysical Journal Letters}, Vol. 955, No. 2, Article L36, pp. 1–7, 2023 (Figure 4). DOI: \href{https://doi.org/10.3847/2041-8213/acf85d}{10.3847/2041-8213/acf85d}. © AAS. Reproduced with permission.}
   \label{fig4}
   \end{figure}

Overall, the quiet Sun is predominantly composed of weak-field internetwork regions \citep{2003A&A...408.1115K} while the strong magnetic fields over $1$\,kG are mainly concentrated in the network regions. The average magnetic field strength in the quiet Sun, measured based on the Hanle effect, is approximately $100$\,G \citep{2012A&A...541A..17S}. The magnetic field in the quiet Sun displays a highly fragmented multi-scale structures, primarily consisting of MBPs  \citep{1984SoPh...94...33M,2010ApJ...723L.169R} and magnetic loop structures \citep{2010ApJ...713.1310I}. These MBPs, representing concentrations of magnetic field, are present in both the network and the internetwork, though within the internetwork they tend to be significantly more isolated and occur much less frequently  \citep{2010ApJ...715L..26S}. High-resolution observations (e.g.Hinode/SOT and SUNRISE/IMaX) have revealed the widespread presence of small-scale magnetic loops with a size of about $100-500$\,km. These loops often emerge from the photosphere in U-shaped or $\Omega$-shaped configurations and decay rapidly (with lifetimes of approximately a few minutes), forming localized bipolar structures \citep{2009ApJ...700.1391M}. Since the weak-field regions occupy a large area in the quiet Sun, they most likely harbor the majority of its magnetic energy \citep{2004Natur.430..326T}. However, the precise contribution of weak fields, the connectivity of magnetic loops, and their heating efficiency \citep{1998Natur.394..152S} remain unresolved challenges \citep{2025ApJ...979..139L}. Three-dimensional magnetic field observations indicate that the quiet Sun magnetic field exhibits a "serpentine topology," where magnetic field lines repeatedly bend and reconnect within the photosphere, forming dynamic clusters of magnetic flux tubes \citep{2023ApJ...955L..36C}. Recent observations from DKIST have revealed complex magnetic topology at subgranular scales in the quiet Sun (Figure \ref{fig4}) and discovered strong magnetic fields with a strength approaching 2 kG within the internetwork, specifically in dark intergranular lanes.

\begin{figure}[!ht]
   \centering
   \includegraphics[width=0.9\textwidth,trim=20 170 20 0,clip]{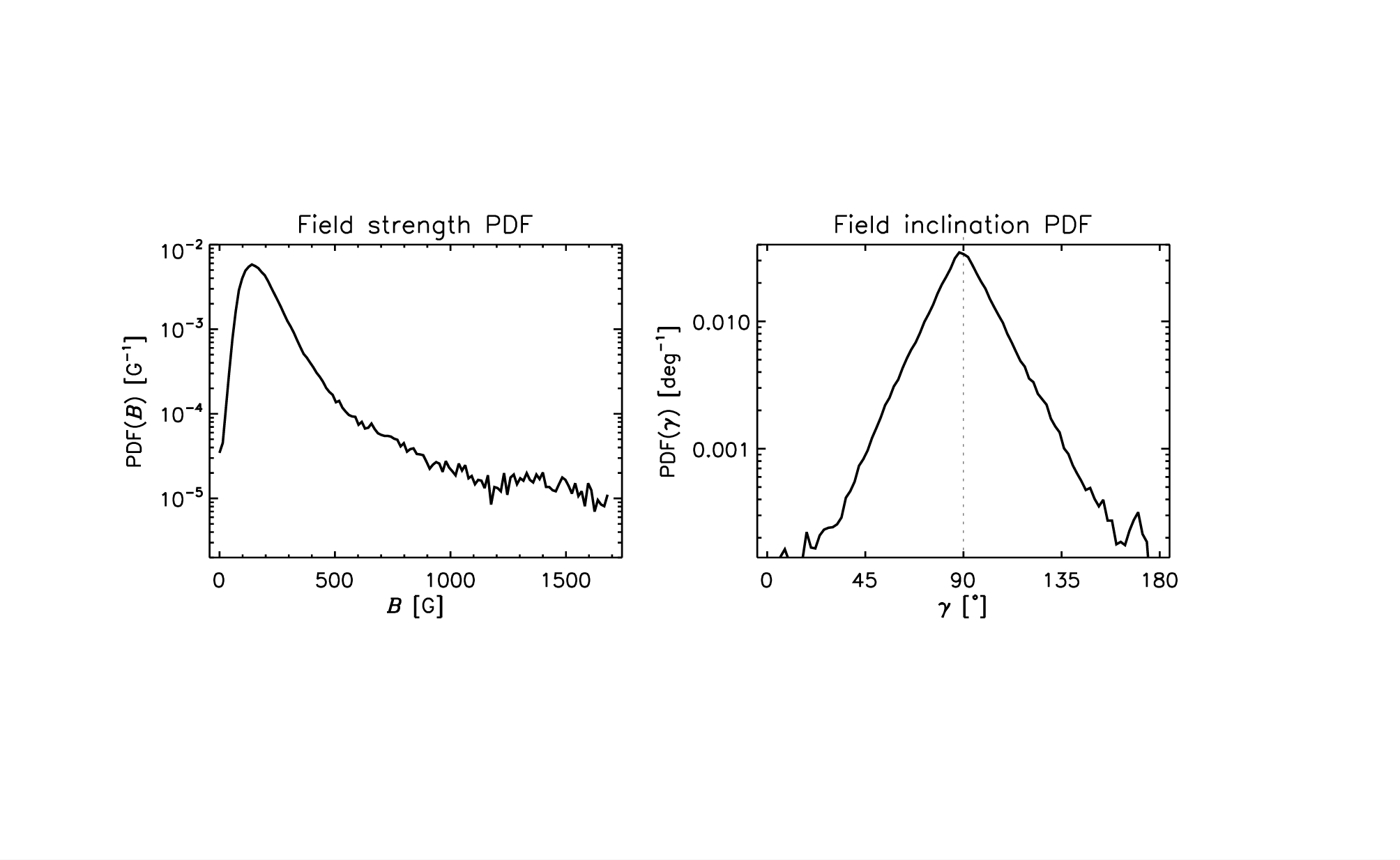}
   \caption{Probability distributions of magnetic field strength and inclination angle in the quiet Sun internetwork obtained from inversion of an ultra-deep Hinode/SP data set, reproduced from \citet{2012ApJ...751....2O}. \newline \small Source: D. Orozco Suárez and L. R. Bellot Rubio, ``Analysis of Quiet-Sun Internetwork Magnetic Fields Based on Linear Polarization Signals,'' \textit{The Astrophysical Journal}, Vol. 751, No. 1, Article 2, pp. 1–13, 2012 (Figure 7). DOI: \href{https://doi.org/10.1088/0004-637X/751/1/2}{10.1088/0004-637X/751/1/2}. © AAS. Reproduced with permission.}
   \label{fig5}
   \end{figure}

The magnetic inclination angle refers to the angle between the magnetic field vector and the line-of-sight direction. Previous statistical studies indicate that the network magnetic field in the quiet Sun tends to be more vertical (with smaller inclination angles) while the relatively weaker internetwork magnetic field tends to be more horizontal \citep{2019LRSP...16....1B}. Figure \ref{fig5} displays results from linear polarization observations (from Hinode/SP) which confirm the presence of a significant horizontal magnetic field component in the quiet Sun internetwork, far exceeding its proportion in active regions \citep{2012ApJ...751....2O}. However, because the internetwork magnetic field is generally weak, resulting in lower signal-to-noise ratio data, it remains inconclusive whether horizontal or vertical magnetic fields dominate within the internetwork. When the data include more MBPs or dark intergranular lanes, the measured vertical field component increases, and the inclination angle decreases \citep{2012A&A...539A...6B}. Nevertheless, there is substantial evidence suggesting a trend of increasing magnetic inclination angle (i.e.the field becoming more horizontal) as the magnetic field strength decreases \citep{2019LRSP...16....1B,2016A&A...593A..93D}.

The analytical results and research studies based on the spectropolarimetric data at disk center from Hinode suggest that the internetwork magnetic field exhibits a quasi-isotropic distribution \citep{2009ApJ...701.1032A}, and such a finding is also supported by \citep{2012ApJ...758L..40M}. However, only a small fraction of the data in these studies have high signal-to-noise ratio polarization signals. Subsequently, Borrero and Kobel analyzed a large set of spectropolarimetric data at various heliocentric angles \citep{2013A&A...550A..98B}. Their results, conversely, indicated that the internetwork field distribution is not isotropic. Other studies have found that the magnetic field is more isotropic near the solar surface, while it becomes more horizontally aligned in the middle to high photosphere \citep{2007ApJ...659L.177H,2008ApJ...672.1237L,2012ApJ...758L..38O}. Whether the quiet Sun internetwork magnetic field is truly isotropic remains a debated topic.

\subsection{Relationship Between Small-Scale Magnetic Flux and the Sunspot Cycle}

The evolutionary behavior of network fields over the solar cycle remains inconclusive. The main viewpoints can be summarized into three categories: no cyclic variation \citep{2002ApJ...564.1042S,2004Natur.430..326T}; anti-correlation with the sunspot cycle \citep{1979ApJ...229L.145G,2003ApJ...584.1107H}; and positive correlation with the sunspot cycle \citep{1973SoPh...28...61H,2003A&A...405.1107M}. These conclusions are often drawn from analyses of intermittent observational data obtained with telescopes of different resolutions, different algorithms and logical reasoning. However, the variation of sunspot numbers over the solar cycle indicates that within an activity cycle, the daily and monthly variations of sunspot numbers do not follow a steady increase or decrease, and they themselves exhibit significant fluctuations. Therefore, it is very easy to lead a wrong solar cycle behavior based on intermittent data. Long-term, continuous, and stable observations are particularly crucial for the study of the solar cycle.

Based on continuous magnetic field observations of Solar Cycle 23 from the space-borne telescope SOHO/MDI \citep{2011ApJ...731...37J,2012ApJ...745...39J}, as well as 15 years of continuous magnetic field observations from SDO/HMI \citep{2026ApJS..283...33J}, it has been found that the correlation between the network magnetic field and the sunspot cycle depends on the flux magnitude of the magnetic structures.  Specifically, the network magnetic structures with magnetic flux greater than 10$^{19}$ Mx vary in phase with the sunspot cycle. The butterfly diagram formed by its latitudinal distribution is about twice as wide as that of the sunspot region, and their hemispheric asymmetry is consistent with that of active regions, which indicate that they originate from the fragmentation and diffusion of active region flux. Conversely, the network magnetic structures with magnetic flux less than 10$^{19}$ Mx constitute majority of the total number of network structures, and their number decreases significantly during solar maximum, showing a clear anti-correlation with the sunspot number.

For the weak internetwork field, most studies point that it has no obvious cyclic variation \citep{2013A&A...555A..33B,2015ApJ...806..174J,2015ApJ...807...70J}. Spectropolarimetric measurements from Hinode/SOT showed that the magnetic flux density in the internetwork region remained constant during solar cycle 24 \citep{2015ApJ...806..174J,2015ApJ...807...70J}.  Furthermore, there is no obvious changes of the latitude-time image of the weak magnetic signals \citep{2014PASJ...66S...4L}. Combined with Hinode observations of horizontal fields \citep{2011ApJ...737...52L} and the power-law distribution of the flux spectrum \citep{2003ApJ...584.1107H}, these evidences strongly suggest the existence of a local small-scale dynamo mechanism, independent of the sunspot dynamo, that continuously generates internetwork magnetic fields. The magnetic structures in internetwork region have typical lifetimes of $5-10$ minutes, and exhibit a highly horizontal orientation \citep{2009ApJ...690..279J}. Although these magnetic elements have low flux, they can transport magnetic flux on the order of 10$^{25}$ Mx per day to the solar surface, constituting a significant energy source for coronal heating \citep{2019LRSP...16....1B}.

\subsection{Diagnostic Techniques for the Quiet Sun Magnetic Field}

Diagnostic techniques for the quiet Sun magnetic field can be divided into indirect and direct methods. The so-called indirect method typically infers the presence of strong magnetic fields from the phenomenon of intensity enhancement exhibited by small-scale magnetic structures in certain wavelength bands. Classic diagnostic bands include H$\alpha$, Ca II H/K lines, the G-band, and the CN-band. However, this qualitative association is only useful for detecting strong, predominantly vertical magnetic fields (>500 G). It can not be used to determine magnetic polarity or true field strength. Currently, the primary method for obtaining full vector magnetic field information remains based on the principle of Zeeman effect. Through the measurement and analysis of polarized radiation information (i.e., the Stokes signals), and under certain assumptions of atmospheric models, the strength and orientation of the magnetic field are derived through an inversion process. This approach is referred to as the direct diagnostic method. In the following section, we will introduce the two main physical effects used in direct diagnostics: the Zeeman effect and the Hanle effect.

The Zeeman effect \citep{2008A&A...477..953M,2002A&A...389..314S,1979ApJ...229..387T,1897ApJ.....5..332Z} is the fundamental physical mechanism for diagnosing solar magnetic fields. When atoms are situated within a magnetic field, their energy levels split, causing spectral lines to split into multiple components with specific polarization states. By precisely measuring the intensity, wavelength shift (Stokes I parameter) circular polarization (Stokes V)and linear polarization (Stokes Q/U) of these split spectral lines, one can quantitatively infer the magnetic field strength, inclination, azimuth, and filling factor \citep{1996ApJ...460.1019L,1973SoPh...32...41S}. For the research studies in quiet Sun, high-sensitivity  high-resolution spectropolarimetry is a core technique. For example, leveraging the properties of spectral lines which are more sensitive to magnetic fields and less affected by non-thermal velocities, researchers have successfully detected the widespread weak and horizontal magnetic field components in the quiet Sun internetwork regions  \citep{2009ApJ...700.1391M,2010A&A...513A...1D}, by using data from space-based (e.g., Hinode/SOT) or balloon-borne (e.g., SUNRISE/IMaX) telescopes.

The Hanle effect is an important and effective diagnostic for weak magnetic fields in the quiet Sun. Its principle is based on the depolarizing effect of magnetic fields on scattered polarized light. When atoms in the solar photosphere are excited by an anisotropic radiation field, producing coherent quantum states, a linear polarization signal is formed. Weak magnetic fields destroy the quantum coherence between atomic energy levels, reducing the degree of linear polarization, with the amount of depolarization related to the magnetic field strength and direction. This characteristic makes it highly sensitive to the weak, often isotropic magnetic fields prevalent in the quiet Sun \citep{2011ApJ...731L..21S}. Diagnosing magnetic fields using the Hanle effect generally involves three steps. The first step is to observe the polarization signals. A high-sensitivity spectropolarimeter is used to measure the linear polarization (Stokes Q/U) of specific spectral lines, such as Fe I 5250.2  {\AA}, which are sensitive to the Hanle effect. The second step is modeling and inversion. By combining radiative transfer equations with atomic physics models, the theoretical polarization degree in the absence of a magnetic field is calculated and then compared with observations. The reduction in polarization degree is directly related to the magnetic field strength, while the polarization direction reflects the azimuth angle of the field \citep{2009ApJ...701.1032A}. The final step is to determine the magnetic field properties. The Hanle effect is highly sensitive to weak fields in the range of $1-100$\,G, and is particularly effective for detecting small-scale horizontal magnetic fields.

The Zeeman effect offers high precision and directness in measuring strong magnetic fields. It can provide the full vector information, namely, the field strength, inclination, and azimuth and, being a mature technique, is widely used in high-resolution magnetogram observations. However, it is limited in its ability to detect weak fields below 10 G. The signal-to-noise ratio is low because the spectral line splitting is small and easily drowned out by thermal motions and instrumental noise, leading to insufficient capability for detecting weak fields in the quiet Sun. Furthermore, the Zeeman inversion technique requires the assistance of complex solar atmospheric model. When the polarization signals are very weak, the inversion method usually faces the problem of non-vonvergence. The advantage of the Hanle effect lies in its ultra-high sensitivity to weak magnetic fields, enabling the detection of magnetic fields as weak as $1-10$\,G. It is particularly adept at capturing the statistical properties of horizontal weak fields in the quiet Sun, without relying on line splitting. However, it generally provides only a probability distribution of magnetic field strength rather than a specific spatial distribution and cannot resolve individual magnetic elements. It also relies on specific spectral lines (e.g.some ultraviolet or molecular lines), requiring high polarization accuracy and precise calibration, which limits its applicability to specific targets. Additionally, the Hanle effect cannot directly measure the longitudinal magnetic field and needs to be combined with Zeeman data for complementarity.

Modern solar magnetic research often combines both effects to fully utilize their complementary advantages. For instance, the SUNRISE mission simultaneously utilized both the Zeeman and Hanle effects to comprehensively characterize the continuous distribution from strong magnetic elements to weak magnetic flux \citep{2011ApJ...731L..21S,2010ApJ...723L.127S}. This synergistic strategy significantly enhances the completeness of global solar magnetic field models and is particularly crucial for understanding the local dynamo and magnetic flux transport mechanisms.

\subsection{Magnetic Energy Transport and Dissipation in the Quiet Sun}

\begin{figure}[!ht]
   \centering
   \includegraphics[width=0.9\textwidth]{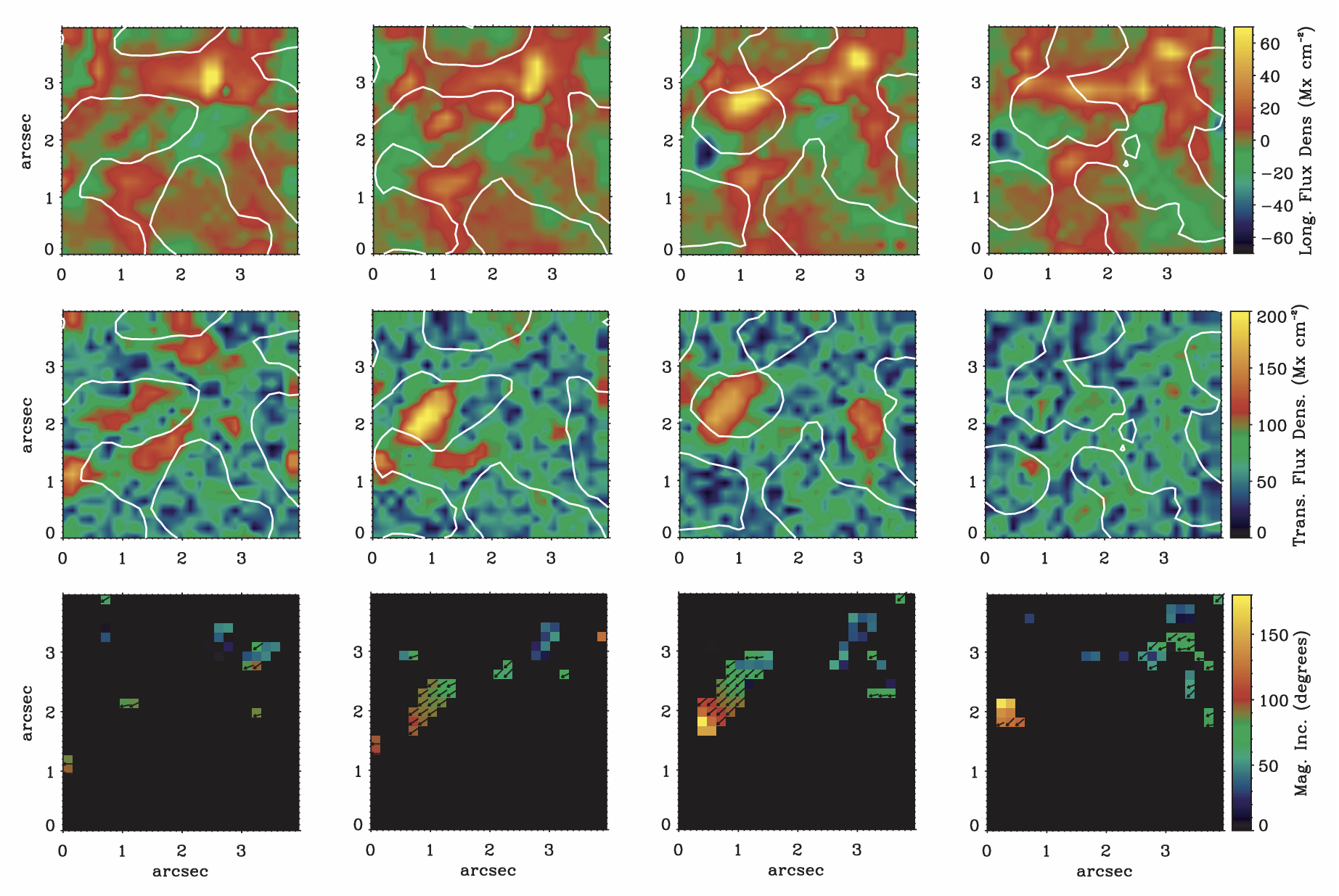}
   \caption{Four sequential time frames (t = 0,125,250,375 s) showing the complete process from emergence to maturation of a magnetic loop, reproduced from \citet{2007ApJ...666L.137C}. Top row: Line-of-sight (longitudinal) magnetic flux density (units:Gauss). The white contours mark the granule-intergranular lane boundaries. Middle row: Transverse magnetic flux density (units:Gauss). The white contours are the same as above. Bottom row: Magnetic field topology (the color scale represents the magnetic field inclination, and the arrows indicate the azimuth angle). The figure shows that the total longitudinal magnetic flux had already reached $10^{17}$\,Mx by $t = 125$ \,s. \newline \small Source: R. Centeno et al., ``Emergence of Small-Scale Magnetic Loops in the Quiet-Sun Internetwork,'' \textit{The Astrophysical Journal Letters}, Vol. 666, No. 2, pp. L137–L140, 2007 (Figure 2). DOI: \href{https://doi.org/10.1086/521726}{10.1086/521726}. © AAS. Reproduced with permission.}
   \label{fig6}
   \end{figure}

The complex and efficient process of small-scale magnetic energy transport in the quiet Sun is crucial for sustaining coronal heating. Magnetic flux within the internetwork is continuously injected into the photosphere primarily through small-scale magnetic emergence events, manifesting as the appearance of $\Omega$-loops or U-loops \citep{2011SoPh..269...13T}. Observations by \citep{2007ApJ...666L.137C} and \citep{2009ApJ...700.1391M} show that these loops rise with velocities of $0.5-2$\,km\,s$^{-1}$ and possess vertical magnetic fluxes of $10^{16}-10^{17}$\,Mx. Figure \ref{fig6} illustrates the complete process, from appearance to maturity, of such a magnetic loop observed by Hinode SP/SOT. The energy flux density injected into the solar atmosphere by small-scale emerging loops in the internetwork alone can reach $2\times{10}^6$\,erg\,cm$^{-2}$\,s$^{-1}$ \citep{2009A&A...495..607I,2010ApJ...714L..94M}. The radiative loss energy flux in the quiet Sun chromosphere is about $4\times10^6-10^7$\,erg\,cm$^{-2}$\,s$^{-1}$. Therefore, these small-scale loops can provide $20\%$ to $50\%$ of the energy required to balance chromospheric radiative losses. This newly emerged magnetic flux rapidly interacts with the vigorous convective motions, undergoing stretching, twisting, fragmentation, and merging \citep{2012ApJ...752..149I,2015ApJ...814..134I,2014ApJ...796...79U,2014ApJ...789....6R,2015ApJ...810...79R}. About $40\%$ internetwork flux ultimately converges and submerges at the suppergranular boundaries \citep{2014ApJ...797...49G}.

Driven and modulated by convection and turbulence, the plasma and magnetic field in the quiet Sun near the solar surface are highly dynamic. During convective processes, magnetic loops can submerge into the solar interior along with the plasma, resulting in the disappearance of magnetic flux from the solar atmosphere \citep{1985MPARp.212..178L, 1988cait.rept.....M}. Although the majority of these loops are confined to the photosphere, approximately $25\%$ can reach the chromosphere and higher layers\citep{2009ApJ...700.1391M}. Horizontal plasma motions at the solar photosphere can bring emerging magnetic loops close to each other or to the background field, leading to magnetic reconnection. This process releases magnetic energy, converting it into thermal and kinetic energy of the plasma \citep{2009RAA.....9..921Z}. Analyzing the time delay between magnetic cancellation events in the photosphere and chromosphere, \citep{1999ASPC..184..222W} proposed that magnetic reconnection occurring just above the photosphere leads to the formation of a U-shaped loop above the reconnection site that lifts upwards, while an $\Omega$-shaped loop formed below sinks into the solar interior. The upward velocity of the U-loop is much greater than the sinking speed of the $\Omega$-loop \citep{1999ASPC..184..222W}. The supersonic upflows discovered by  \citep{2010ApJ...723L.144B} likely correspond to reconnection outflows generated in similar magnetic cancellation processes. The observed "serpentine topology" of the quiet Sun magnetic field is also often attributed to changes in magnetic connectivity caused by reconnection between magnetic loops \citep{2023ApJ...955L..36C}. Different types of magnetic reconnection may occur at various heights from the photosphere to the low corona, manifesting as small-scale magnetic activities observed in different wavelengths, such as the H$\alpha$ line wings (quiet sun Ellerman bombs) \citep{2013ApJ...779..125N, 2017ApJ...845...16N}, ultraviolet lines \citep{2014ApJ...797...88H}, and extreme ultraviolet lines \citep{2021A&A...656L...4B,2021A&A...656L...7C}.

Convective and turbulent motions can also induce oscillations in the plasma and magnetic field, exciting various types of magnetohydrodynamic (MHD) waves \citep{2007Sci...318.1572E,2019AGUFMSM13B..09L,2011Natur.475..477M,2022SoPh..297...20S}. Upwardly propagating MHD waves carry energy into the chromosphere and corona. During their propagation, they undergo mode conversion and dissipation, thereby heating the solar atmosphere \citep{2014masu.book.....P}. High-resolution observations have revealed fine-structured current sheets within and around MBPs \citep{2023ApJ...955L..36C,2010ApJ...723L.164L,2010ApJ...723L.127S}. These sites are crucial for small-scale magnetic reconnection, local heating, and wave excitation, and serve as important source regions for chromospheric jets, spicules, and the solar wind \citep{2024NatAs...8.1246H, 2014ApJ...797L..14T}. Theoretical models and observations indicate that the energy stored and transported by the small-scale magnetic field in the quiet Sun is sufficient to explain a large portion of the heating requirements in the chromosphere and transition region \citep{2011ApJ...731L..21S}. Therefore, the energy transport process associated with the quiet Sun magnetic field is a complex and highly efficient dynamic process: powered by the local dynamo, driven by small-scale magnetic emergence, organized by convective and turbulent flows, with the energy ultimately released through magnetic reconnection and wave dissipation \citep{2015Natur.522..188A}.


\section{Theory and Numerical Studies of Small-Scale Turbulent Dynamo}

High-resolution observations from facilities such as Hinode, SUNRISE, SST, GREGOR, GST, and DKIST have provided abundant evidence for the prevalence, fine-scale topology, and dynamic evolution of small-scale magnetic fields in the quiet Sun. These evidences strongly suggest the existence of a local turbulent dynamo. However, to gain a deep understanding on how magnetic fields are generated, amplified, maintained, and shaped into the observed small-scale magnetic features in the quiet Sun through turbulence, existing observational data alone are far from sufficient. Theoretical studies of the small-scale solar dynamo and the use of numerical simulation tools are crucial for uncovering its underlying principles. A synergistic approach combining theory, numerical simulations, and observations is essential for effectively elucidating the intrinsic operational mechanisms of the small-scale solar dynamo. This section will systematically elaborate on the core physics of turbulent dynamo, sparticularly the small-scale dynamo, review key developments in its theoretical models, and highlight major breakthroughs achieved through magnetohydrodynamic (MHD) numerical simulations in reproducing the observed characteristics of the Sun's small-scale magnetic field and probing its deep-seated origins.

\subsection{Theoretical Foundation and Principles}

The fundamental concept of SSD dates back to the 1950s \citep{1950QJRMS..76..133B}. In recent decades, advancements in computational hardware and software have enabled scientists to gain a deeper understanding of the SSD through magnetohydrodynamic (MHD) numerical simulations.

The SSD describes the process within a turbulent system where magnetic field fluctuations on spatial scales smaller than the turbulent energy injection scale (e.g., the convective cell scale) are continuously and rapidly amplified, relying on random turbulent motions. The underlying physical principle is that turbulent motions stretch, twist, and fold the magnetic field, converting kinetic energy into magnetic energy, thereby amplifying the field by orders of magnitude. From the base of the convection zone (CZ) to the solar surface, convective cells gradually decrease in size and rapidly reverse \citep{1953ZA.....32..135V}. Convective motions drive turbulence, and when this turbulence is sufficiently strong, it can amplify the magnetic field on the scale of convective cells or smaller. Whether driving turbulence with different forms and mechanisms in the system has a significant impact on the properties and formation characteristics of the small-scale dynamo is still highly controversial \citep{2011ApJ...736...36M}.

The principle of the turbulent dynamo can be described simply by the magnetic induction equation:           
\begin{equation}
    \frac{\partial \bf{B}}{\partial t}=\mathrm{\nabla}\times(\bf{v} \times \bf{B})+\eta\mathrm{\nabla}^2\bf{B}
\end{equation}
Here, $\mathbf{B}$ is the magnetic field, and $\eta$ is the magnetic diffusivity. This equation shows that the temporal evolution of the magnetic field depends crucially on the two terms on the right-hand side: the core term ${\nabla}\times(\bf{v} \times \bf{B})$ can cause the amplification of magnetic field due to the stretching of magnetic field lines by turbulence, while $\eta \nabla^2\bf{B}$ is the diffusion (dissipation) term. A net increase in magnetic energy over time requires that the turbulent amplification term exceeds the magnetic diffusion term. To fully understand how turbulence in the system affects plasma velocity evolution in space and time and consequently influences the magnetic field, as well as the temporal evolution of magnetic, kinetic, and thermal energy, it is generally necessary to solve this equation coupled with the equation of motion and the energy equation.

The magnetic Prandtl number ($P_m$) is another crucial dimensionless parameter in the small-scale dynamo process, defined as $P_m = R_m / R_e$, where $R_m$ is the magnetic Reynolds number and $R_e$ is the fluid Reynolds number. The magnetic Reynolds number $R_m = |{\nabla}\times(\bf{v} \times \bf{B})|/|\eta \nabla^2\bf{B}|\approx L U/\eta$,where $L$ is the characteristic inertial scale of the system and $U$ is the typical characteristic velocity at the largest scale. The corresponding fluid Reynolds number is defined as $R_e = LU/\nu$, where $\nu$ is the kinematic viscosity. Thus, the magnetic Prandtl number can also be simplified to $P_m =\nu/\eta$.

Generally, magnetic field amplification on small scales occurs more readily when $P_m\gg1$. From the base of the solar convection zone to the photosphere, $P_m$ ranges from approximately $10^{-6}$ to $10^{-4}$ \citep{2020RvMP...92d1001S}. This means that in this region, the magnetic diffusivity ($\eta$) is several orders of magnitude larger than the kinematic viscosity ($\nu$) and the dissipation scale of fluid motions is much smaller than the magnetic diffusion scale. Furthermore, for turbulence to be generated and initiate the dynamo process in a system, the magnetic Reynolds number ($R_m$) must exceed a critical value, $R_m^{crit}$. Existing MHD simulation results indicate that $R_m^{crit}$ varies with $P_m$ (see details in section 3.4), and $R_m^{crit}$ is relatively larger for smaller $P_m$. Simulation results indicate that $R_m^{crit}$ is close to or less than 400 \citep{2012PhyS...86a8404K}. Based on the magnetic diffusivity $\eta$ with Spitzer formular \citep{2020RSPSA.47690867N,2022A&A...665A.116N,1962pfig.book.....S}, the value of $\eta$  at the solar surface is approximately $10^4$\,m$^2$\,s$^{-1}$\citep{2020RSPSA.47690867N}. With a typical surface velocity of about $1000$ m\,s$^{-1}$ and a characteristic length scale of $1$\,Mm (a typical granule size), the resulting magnetic Reynolds number is approximately $10^5$. This value far exceeds the critical $R_m^{crit}$ derived from numerical simulations, indicating that turbulence can readily develop near the solar surface. However, in numerical studies, a sufficient number of grid points is typically required to ensure that the numerical diffusion is much smaller than the physical diffusion. A smaller $P_m$ implies that numerical computations require more grid points to resolve down to the fluid motion dissipation scale. Achieving MHD simulations of the small-scale turbulent dynamo under the realistic $P_m$ conditions of the solar photosphere and convection zone demands extremely high spatial resolution, which currently remains a formidable challenge.

\begin{figure}[!ht]
   \centering
   \includegraphics[width=0.9\textwidth]{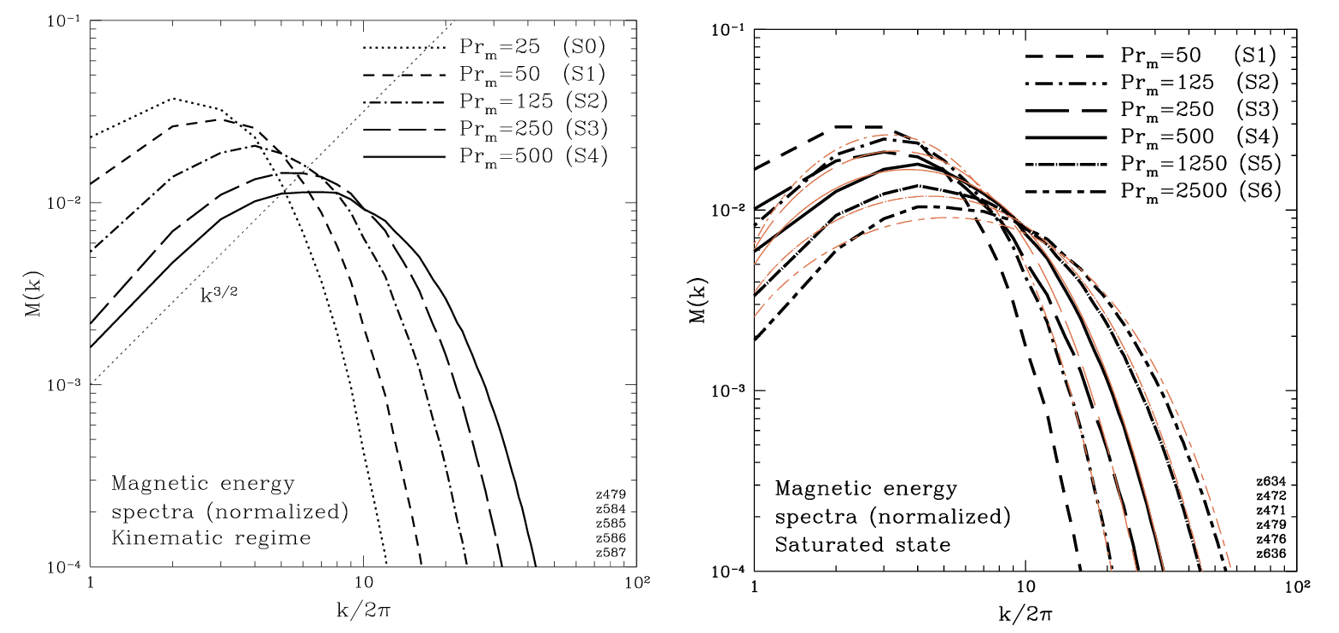}
   \caption{Distribution of the system's normalized magnetic energy M(k) versus wavenumber k for different $P_m$ values during the kinematic stage (left) and nonlinear saturation stage (right), reproduced from \citet{2004ApJ...612..276S}. \newline \small Source: A. A. Schekochihin et al., ``Simulations of the Small-Scale Turbulent Dynamo,'' \textit{The Astrophysical Journal}, Vol. 612, No. 1, pp. 276–307, 2004 (Figures 4(b) and 20). DOI: \href{https://doi.org/10.1086/422547}{10.1086/422547}. © AAS. Reproduced with permission.} 
   \label{fig7}
   \end{figure}

The growth process of the small-scale turbulent dynamo is divided into the kinematic stage and the nonlinear saturation stage. Figure \ref{fig7} from \citep{2004Ap&SS.292..141S} shows the magnetic energy $M(k)$ distribution versus wavenumber $k$ at different $P_m$ values during both the kinematic and nonlinear saturation stages. During the kinematic stage, turbulent vortices drive exponential growth of magnetic energy through random stretching of the magnetic field, forming characteristic folded structures. In these structures, the magnetic field direction reverses frequently at the resistive scale (magnetic diffusion scale), while field line curvature occurs at the flow scale \citep{2004ApJ...612..276S}. The magnetic energy growth rate $\gamma$ is given by $\gamma \propto u_d/l_d$ \citep{1968JETP...26.1031K}, where $l_d$ and $u_d$ are the dissipation length scale and velocity, respectively. By mapping an appropriate turbulence model (e.g., Kolmogorov model) onto the Kazantsev theory, one can substitute $u_d$ and $l_d$ to eventually estimate the growth rate $\gamma$. The growth rate $\gamma$ depends on both $P_m$ and $R_m$. Near the critical magnetic Reynolds number $R_m^{crit}$, numerical simulation results indicate that $\gamma$ is proportional to $ln(R_m/R_m^{crit})$ in the small Pm range \citep{2023NatAs...7..662W,2026arXiv260401718K}. However, this relationship may not represent the actual scaling law of the growth rate in environments with significantly larger $R_m$ values. 

Based on the Kazaztsev theory, the relationship between the magnetic energy spectrum and the wavenumber can be derived. During the kinematic stage of dynamo growth, when magnetic fluctuations are still relatively weak, the generated magnetic structures have sizes comparable to the resistive scale, but the field lines are curved on the vortex scale. Consequently, the peak of the magnetic energy spectrum is expected to occur at the resistive scale. For scales larger than the resistive scale, the magnetic energy exhibits a power-law spectrum of the form $E(k) \sim k^{3/2}$. As the scale decreases and the wavenumber increases, the magnetic energy eventually drops abruptly and sharply as shown in Figure. 7 \citep{2005PhR...417....1B,2019JPlPh..85d2001R}. Numerical simulation results further indicate that the spectral index of magnetic energy varies with $P_m$, deviating from the $3/2$ exponent (see Section 3.2 for details). Upon entering the nonlinear saturation stage, fluid motions on scales smaller than the resistive scale are suppressed by the Lorentz force. The magnetic energy reaches its peak and no longer increases with time. Simulations with high $P_m$ values \citep{2004ApJ...612..276S} show that stretching along the magnetic field direction is inhibited, while quasi-two-dimensional mixing in the perpendicular direction is enhanced, leading to the cessation of magnetic energy growth. The folded structures persist but undergo directional reversals, and the magnetic energy spectrum becomes flatter compared to the kinematic stage. In both high and low $P_m$ plasma environments, the final magnetic energy remains less than the kinetic energy of the system. Numerical simulations of the solar small-scale turbulent dynamo in the nonlinear saturation stage often yield magnetic field distributions in the quiet Sun that are consistent with observational results.

In addition to SSD operating on scales below convective cells, large-scale dynamos (LSD) also exist in the convective zones of the Sun and stars. The physical principle of the LSD involves the generation of an $\alpha$-effect by helical turbulence (where $\langle \bf{v} \cdot \nabla \times \bf{v} \rangle \neq 0$) to produce a mean electric field $\bf{E} = \alpha\langle\bf{B}\rangle$. This couples with the shear from differential rotation (the $\Omega$-effect) to excite periodic magnetic fields, collectively driving the evolution of the large-scale magnetic field \citep{2012SSRv..169..123B}. Unlike the small-scale dynamo, the large-scale magnetic field is amplified through the statistical mean field, rather than relying on local, random turbulent fluctuations \citep{2005PhR...417....1B}. There are two key driving factors. One is the tachocline, for example, the angular velocity difference between the Sun's equator and polar regions which provides energy for the stretching and amplification of the large-scale magnetic field \citep{2014ApJ...785...49P}. This is also why large-scale turbulent dynamo action tends to occur at the base of the convection zone \citep{2020LRSP...17....4C}. The other is helical turbulence: the helical structures in turbulence induce large-scale currents through the $\alpha$-effect, resulting in a periodic conversion between poloidal and toroidal magnetic fields \citep{2005PhR...417....1B}.
 
The magnetic fields generated by the LSD are spatially more extensive and can operate even under conditions of low magnetic Prandtl number ($P_m \ll 1$)\citep{2005ApJ...625L.115S}. The global solar dynamo model is a typical representative of the LSD, generating the 11-year magnetic cycle through the interaction between the tachocline (a layer of strong shear) and the turbulent convection zone \citep{2012SSRv..169..123B}.Notably, among various large-scale dynamo models, the flux transport dynamo performs excellently in explaining observations \citep{2023SSRv..219...39H}. Based on the traditional $\alpha-\Omega$ framework, this model introduces the meridional circulation and the Babcock - Leighton mechanism, and can successfully reproduce the butterfly diagram of sunspots, the poleward propagation of the polar field, and the irregularities of the solar cycle (such as the Waldmeier effect and the Maunder minimum). In galactic magnetic field evolution, the coupling of large-scale shear with turbulence may explain the formation of ordered magnetic structures \citep{2005PhR...417....1B}.

In most regions of the solar convection zone, instabilities associated with these two distinct types of turbulent dynamos are often excited simultaneously and nonlinearly influence each other.

\subsection{Numerical Methods for studying Turbulent Dynamos}

Solving the magnetohydrodynamic (MHD) equations is a crucial method and approach for studying the small-scale turbulent dynamo. When the computational grid scale is smaller than the dissipation scales, such as those for magnetic diffusion and viscous dissipation, the solution process can directly capture the physical processes occurring at these dissipation scales. This approach is referred to as Direct Numerical Simulation (DNS). However, DNS is not particularly suitable for simulating turbulent dynamos in solar or large-scale astrophysical environments. At the solar surface, assuming the typical characteristic velocity of $1$\,km\,s$^{-1}$, the typical magnetic diffusivity $\eta$ of approximately $10^4$\,m$^2$\,s$^{-1}$, the typical kinematic viscosity $\nu$ of about $0.1$\,m$^2$\,s$^{-1}$ and a characteristic length $L$ of $1$\,Mm (the scale of a convective cell), then the magnetic diffusion scale is calculated as $l_\eta \sim L\ Rm^{-3/4}\approx200$\,m, while the viscous dissipation scale is $l_\nu~\sim L\ R_e^{-3/4}\approx0.03$\,m. The magnetic diffusivity and kinematic viscosity due to collisions among ions, electrons, and neutrals can be calculated in detail following the previous papers \citep{2020RSPSA.47690867N,2022A&A...665A.116N,2020RvMP...92d1001S}. If the simulation domain spans the characteristic scale of $1$\,Mm, then using the DNS to resolve down to the true viscous dissipation scale of $0.03$\,m would require about $10^7$ grid points in each direction. Such a massive computation and data processing capability is currently not feasible.

The Large Eddy Simulation (LES) method is commonly employed in turbulent dynamo studies. The core premise of LES is that large-scale eddies dominate turbulent transport and energy distribution. Therefore, if a numerical simulation can explicitly resolve these large-scale structures and appropriately handle the effects of unresolved small scales, it can provide a realistic description of the flow for practical applications. The effects of unresolved small scales are typically incorporated using either explicit subgrid-scale (SGS) models or implicit numerical dissipation.

In Explicit Large Eddy Simulation (ELES), the solved equations are spatially filtered at a scale larger than the grid scale, and (ideally) a physically motivated SGS model is employed to describe the terms corresponding to the scales that are filtered out during the filtering process. A famous example is the Smagorinsky scheme \citep{1963MWRv...91...99S} which introduced the concept of turbulent eddy viscosity as an SGS model. Subsequently, the ELES method was gradually refined and developed, with more physical terms beyond viscosity being incorporated through subgrid-scale approaches  \citep{2014arXiv1412.2740C,2017PhRvE..95c3206G}. The ELES method has been used to study decaying MHD turbulence, where its higher computational efficiency compared to DNS has been presented and explained in detail. It has also been applied to the studies of the local interstellar medium \citep{2008ApJ...686.1137C} and the kurtosis and flatness of turbulent flows \citep{2009essu.confE..75C}. ELES is still widely used in large-scale simulations of convection-driven dynamos \citep{2008ApJ...676..680F,2009LRSP....6....4F,2014ApJ...789...35F}, but it is less commonly applied in small-scale dynamos.

Implicit Large Eddy Simulation (ILES) is a type of LES simulation that does not include any explicit SGS model. In this approach, the physical dissipation terms are replaced by the inherent dissipation and dispersion arising from the nature and characteristics of the numerical algorithm \citep{2011PhFl...23c4106G,2015LRCA....1....2S}. The primary advantage of this method lies in its high effective resolution; dissipation only occurs at the grid scale, so that larger scales are almost unaffected by artificial diffusion. However, when using ILES simulations, the actual values of the dimensionless control parameters become ill-defined, making it difficult to assess the impact of the dissipation scheme on the resolved plasma flow. This method has various specific implementations, ranging from hyperviscosity operations to slope-limited dissipation schemes, among others. The ILES method has achieved notable success in simulating near-surface solar dynamos that include radiative transfer processes \citep{2014ApJ...789..132R,1998ApJ...499..914S}. {However, there are significant differences between the DNS-type simulation results and the ILES-type simulation results that include the bottom region of the convection zone \citep{2016Sci...351.1427H} , indicating that the ILES method may not necessarily be suitable for convective systems in which both large-scale and small-scale turbulent dynamos coexist. Nevertheless, these ILES-type numerical simulations provide very meaningful insights into, among other things, the interaction between SSD and LSD.

In simulations of turbulence-driven dynamos, some studies solve the dimensionless MHD equations directly, without including physical effects like gravitational stratification or radiative transfer, and they typically drive turbulence by applying artificial forcing, often using periodic boundary conditions in all directions \citep{2004ApJ...612..276S,2023NatAs...7..662W}. Implementing such simulations is relatively easier. While they can effectively study the fundamental physics of turbulent dynamos in both linear and nonlinear stages, they generally fail to successfully reproduce the distribution and structural characteristics of solar and stellar magnetic fields. In contrast, dynamo simulations that include radiative transfer and gravitational stratification, and drive turbulence through convection, can reproduce the characteristics of observed magnetic field distributions quite well in the nonlinear stage \citep{2014ApJ...789..132R,2007A&A...465L..43V}.

 \cite{2014ApJ...789..132R} simulated the small-scale dynamo process at different resolutions, they found that the resolution can significantly impact the magnetic energy growth rate during the kinematic stage: smaller grid scales lead to higher growth rates. As resolution increases, the energy spectrum converges at larger scales, while the small-scale part exhibits more significant extension. However, their results also indicated that, at least within the employed ILES numerical framework, the simulation results approach an asymptotic limit on scales above approximately 50 km; increasing resolution only smoothly extends the energy spectrum to smaller scales without major qualitative changes, as shown in the left panel of Figure 8. 
 
Setting appropriate bottom boundary conditions is a common challenge in near-surface solar turbulent dynamo simulations, and different boundary conditions can significantly affect the results. In the work of \cite{2014ApJ...789..132R}, numerous simulations with varying boundary conditions and resolutions were performed. The results simulated under the OSb bottom boundary conditions were closest to the actual conditions on the solar surface. OSb is one kind of open boundaries. Moreover, all three components of the magnetic field and the mass flux components are symmetric at the bottom boundary. Gas pressure is decomposed into mean pressure and fluctuating pressure. The mean pressure is extrapolated into the ghost cells, fixing its value at the boundary, while the pressure fluctuations are damped/dissipated in the ghost cells. As shown in the right panel of Figure \ref{fig8}, the choice of different boundary conditions and resolutions can substantially influence various aspects of the simulation outcomes. 

\begin{figure}[!ht]
   \centering
   \includegraphics[width=0.9\textwidth]{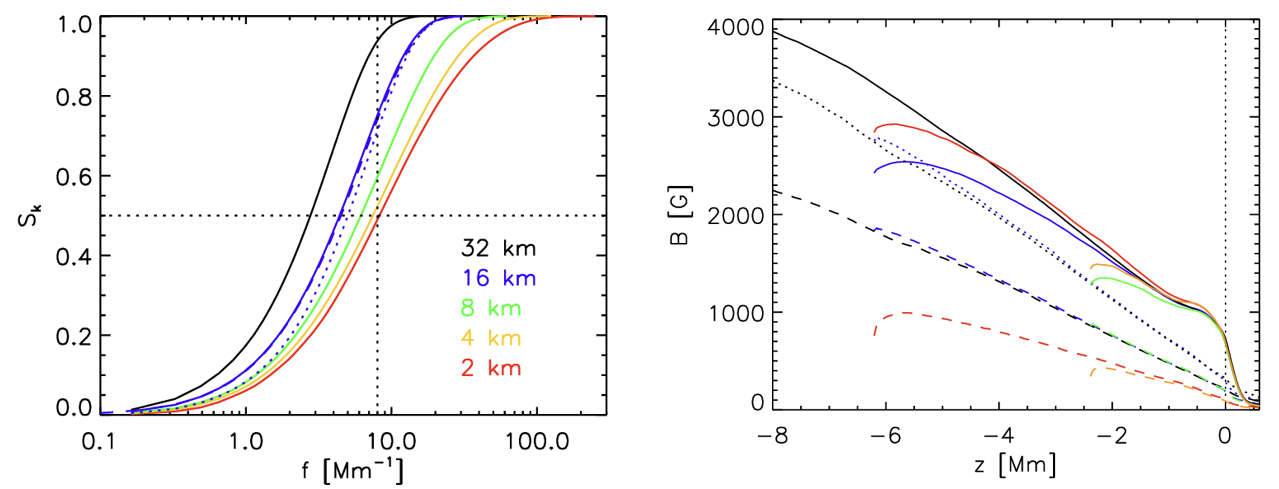}
   \caption{Left panel: Normalized magnetic energy spectra at different spatial resolutions, reproduced from \citet{2014ApJ...789..132R}. When the grid spacing is 2 km, approximately 50\% of the magnetic energy at the $\tau = 1$ level in the photosphere resides on scales smaller than 100 km. Furthermore, smaller grid spacings lead to magnetic energy being more concentrated at smaller scales. Results converge towards the real solar surface conditions only when the grid spacing is less than 8 km. The solid blue, dashed blue, and dotted blue lines represent results under different boundary conditions. Right panel: The height dependence of the root mean square (rms) magnetic field (solid lines) and the equivalent magnetic field strength (dashed lines) for different boundary conditions, all with a grid spacing of 16 km. Lines of different colors represent different boundary conditions, with the blue and green lines corresponding to the OSb boundary condition.\newline \small Source: M. Rempel, ``Numerical Simulations of Quiet Sun Magnetism: On the Contribution from a Small-Scale Dynamo,'' \textit{The Astrophysical Journal}, Vol. 789, No. 2, Article 132, pp. 1–22, 2014 (Figures 4(a) and 11(a)). DOI: \href{https://doi.org/10.1088/0004-637X/789/2/132}{10.1088/0004-637X/789/2/132}. © AAS. Reproduced with permission.}
   \label{fig8}
   \end{figure}

\subsection{Progress in Numerical Simulations of the Small-Scale Turbulent Dynamo at Different $P_m$ Values}

Over the past few decades, significant progress has been made in the numerical simulation of SSD particularly through a large number of studies with the value of $P_m$ close to or greater than $1$. For incompressible, homogeneous, isotropic, and non-helical setups, the previous numerical simulation has confirmed that an SSD can indeed be generated at $P_m=1$ \citep{1981PhRvL..47.1060M}. Then, the subsequent studies have advanced our understanding of the SSD under various configurations in high-$P_m$ regimes \citep{2005PhR...417....1B}. Schekochihin and colleagues conducted DNS-type incompressible MHD simulations with $P_m \geq 1$, and they found that small-scale magnetic fields begin to grow spontaneously at a critical magnetic Reynolds number $R_m^{crit} \approx 60$. Then, the saturated fields exhibit a folded structure and suppress turbulent kinetic energy. The magnetic fields grow exponentially through the stretch-fold mechanism, providing direct supports for theoretical models \citep{2004Ap&SS.292..141S,2004ApJ...612..276S}. They later found that, at achievable resolutions, $R_m^{crit}$ increases with the fluid Reynolds number ($R_e$) without showing signs of approaching a constant limit \citep{2005ApJ...625L.115S}. However, the recent numerical results by solving the Kazantsev model suggest that $R_m^{crit}$ saturates at a constant value of $300$ for $R_e > 10^{5}$ \citep{2026arXiv260401718K}.

The magnetic Prandtl number ($P_m$) in the convective zones of the Sun and stars is typically much less than $1$, while the Reynolds number and magnetic Reynolds number are very large. Simulating the SSD under realistic $R_e$ and $P_m$ values is extremely challenging, even for idealized MHD models. Based on the DNS-type simulations, for the first time, \citep{2007PhRvL..98t8501I} confirmed the existence of the SSD in randomly forced turbulence with $P_m \ll 1$.  \citep{2007NJPh....9..300S} systematically established the theoretical framework for the fluctuation dynamo at low $P_m$, emphasizing the crucial role of inertial-range turbulence in magnetic field amplification. Recently, a series of idealized MHD simulations across different $P_m$ values have been performed, reaching a minimum $P_m$ of 0.0025 \citep{2023NatAs...7..662W}. This is currently the simulation that most closely approaches the $P_m$ value of the actual solar convection zone.  The authors found that SSD is most difficult to generate in the range the range $0.04 < P_m < 0.1$, but becomes easier again for $P_m < 0.04$. Crucially, across all $P_m$ ranges, the SSD operates whenever Rm exceeds a critical value $R_m^{crit}$. For high-$P_m$ plasmas, $R_m^{crit}$ lies approximately between $30-60$ \citep{2005PhR...417....1B}, whereas for low-$P_m$ plasmas, $R_m^{crit}$ is about $200-400$ \citep{2012PhyS...86a8404K,2023NatAs...7..662W}. For $P_m$ < 0.04, $R_m^{crit}$ decreases again to below $100$. It is important to note that these discussions of $R_m^{crit}$ primarily stem from idealized MHD simulation studies.

\cite{2007NJPh....9..300S} found that for $P_m$ values in the range of $0.1-1$, the magnetic field growth rate $\gamma$ decreases monotonically as $P_m$ decreases. For even smaller $P_m$ values, the growth rate gradually approaches a constant. As $P_m$ drops to very low values ($P_m \ll 0.1$), $\gamma$ no longer varies with $P_m$, but gradually converges to a stable constant. Based on this phenomenon, the authors inferred that there is a positive asymptotic growth rate in the limit of high $R_m$ ($R_m \gg 1$) and very low $P_m$. However, simulations at even lower $P_m$ have not yet unambiguously confirmed such a positive asymptotic value. For $P_m > 1$, numerical results show a growth rate scaling of $\gamma\sim{\rm R_m}^{1/2}$, that is consistent with expectations for Kolmogorov turbulence\citep{2005PhR...417....1B}. However, this scaling breaks down in low-$P_m$ simulations \citep{2007PhRvL..98t8501I,2007NJPh....9..300S,2023NatAs...7..662W}. For $P_m < 1$ and $R_m$ close to $R_m^{crit}$, the growth rate follows $\gamma \sim ln(R_m/R_m^{crit})$ \citep{2023NatAs...7..662W,2026arXiv260401718K}, but this may not represent the true scaling law in environments with significantly larger $R_m$.

In incompressible, homogeneous systems, the kinetic energy spectrum in the inertial range follows the relation $E_{kin}(k)~k^{-5/3}$. This spectral relationship is known as the Kolmogorov law of the energy cascade. As shown in Figure \ref{fig9}, for an ideal Kolmogorov spectrum, the kinetic energy spectrum should appear as a straight line within the inertial range, before dropping steeply upon approaching the viscous dissipation scale. However, simulations of the small-scale turbulent dynamo reveal the emergence of a 'bump' structure in the kinetic energy spectrum just before this steep drop-off near the viscous dissipation scale (see the inset in Figure \ref{fig9}). The low-wavenumber side of this 'bump' exhibits a positive deviation from the Kolmogorov scaling and overlaps with the wavenumber associated with the energy-containing scale of magnetic fluctuations. This feature, identified as the bottleneck region in the kinetic energy spectrum, is believed to reduce the efficiency of the SSD \citep{2011ApJ...741...92B,2023NatAs...7..662W}. As mentioned previously, ignoring the low-wavenumber range, the magnetic energy spectrum for scales larger than the resistive scale theoretically follows a power-law form of $E(k) \sim k^{3/2}$. However, the low $P_m$ simulation results indicate that the magnetic energy spectrum does not conform well to $E(k) \sim k^{3/2}$. Furthermore, in the high wave number range, the magnetic energy spectrum exhibits a negative power-law cascade with varying degrees of steepness. In the low magnetic Reynolds number regime, the negative power-law spectral index is $-11/3$, while in the high magnetic Reynolds number regime, it is $-1$ \citep{2023NatAs...7..662W}.

\begin{figure}[!ht]
   \centering
   \includegraphics[width=0.9\textwidth]{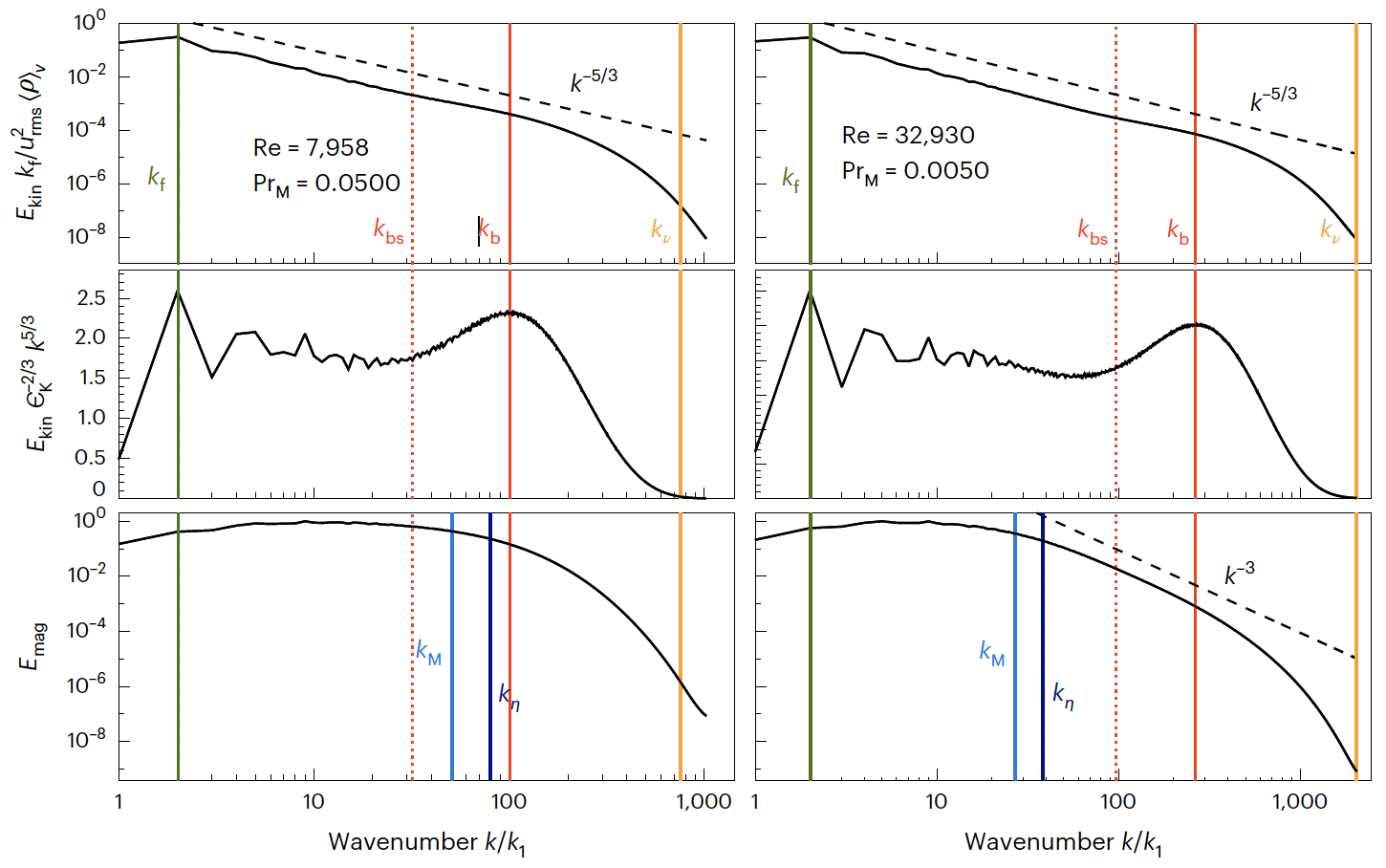}
   \caption{Distribution of kinetic (top two rows) and magnetic energy spectra (bottom row) for two different combinations of Reynolds number ($R_e$) and magnetic Prandtl number ($P_m$), reproduced from \citet{2023NatAs...7..662W}. The middle row shows the kinetic energy spectrum multiplied by $k^{5/3}$ which clearly reveals the bump structure. The vertical solid green line corresponds to the forcing wavenumber $k_f$. The vertical solid red line indicates the wavenumber $k_b$ at the peak of the bottleneck. The vertical dotted red line marks the wavenumber $k_{bs}$ where the bottleneck begins. The vertical solid orange line represents the viscous dissipation wavenumber $k_v$. The vertical solid dark blue line indicates the Ohmic dissipation wavenumber, while the vertical solid light blue line corresponds to the characteristic magnetic wavenumber $k_M$.\newline \small Source: J. Warnecke et al., ``Numerical evidence for a small-scale dynamo approaching solar magnetic Prandtl numbers,'' \textit{Nature Astronomy}, Vol. 7, No. 6, pp. 662–668, 2023 (Figure 4). DOI: \href{https://doi.org/10.1038/s41550-023-01975-1}{10.1038/s41550-023-01975-1}. © AAS. Reproduced with permission.}
   \label{fig9}
   \end{figure}

The above studies are all based on the results of numerical simulations, generated by ideal MHD and random noise-driven turbulence, in the kinematic dynamo stage. Most research on the nonlinear saturation stage of SSD has focused on cases with $P_m \geq 1$, with only a few investigations extending into the $P_m < 1$ regime. For instance, a quasi-DNS approach was applied to study the turbulent dynamo in its nonlinear saturation stage by \citep{2011ApJ...741...92B}.They first ran a case with $P_m = 1$ and let it reach the saturation stage. Then they continuously reduced the kinematic viscous dissipation while keeping the magnetic Reynolds number $R_m$ nearly constant. Subsequently, they adopted the new parameter values and continued the integration from the saturated state. It was found that as the magnetic field strength increases, nonlinear effects become significant, including the back-reaction of the Lorentz force on the fluid motions. These effects can suppress further magnetic field growth, leading to an equilibrium state \citep{2011ApJ...741...92B}. Compared to the $P_m=1$ case, the magnetic field in the $P_m < 1$ regime exhibits a significantly higher filling factor, weaker correlation with the turbulent velocity field, less folding, and generally thicker structures. A key finding was that most of the energy is dissipated via Joule heating before reaching the viscous dissipation scale, thereby allowing the kinematic viscosity to be further reduced compared to the estimated value at a given grid resolution. Another important result is that the saturation level of the SSD shows only a weak dependence on $P_m$. When the $P_m$ value has changed by two orders of magnitude, the saturation strength of the SSD only decreases from approximately $40\%$ of the turbulent equipartition level to nearly $10\%$. Regardless of the scale, the magnetic energy remains less than the kinetic energy. Interestingly, the bottleneck effect observed in the kinematic stage was found to be suppressed in the nonlinear regime.

Simulations of convection-driven small-scale dynamos began to emerge in the 1990s. For the first time, \citep{1992ApJ...392..647N}  obtained the self-sustaining magnetic fluctuations driven by convection \citep{1992ApJ...392..647N}. Subsequently, \citep{1999ApJ...515L..39C} confirmed the existence of the SSD using a Boussinesq approximation setup. Then, under compressible and more realistic gravitational stratification, the solar turbulent dynamo was successfully reproduced, and the observational features of the quiet Sun magnetic fields were obtained \citep{2010A&A...513A...1D,2015ApJ...803...42H,2015ApJ...809...84K,2010ApJ...714.1606P,2014ApJ...789..132R}. In these simulations, the $P_m$ value was typically around or greater than $1$. Currently, Implicit Large Eddy Simulation (ILES) of the near-surface solar SSD have achieved effective $P_m$ values close to $0.1$ \citep{2019ApJ...879...57B,2018ApJ...859..161R}. This was accomplished by employing a more diffusive numerical scheme for the magnetic induction equation together with a less diffusive scheme for the momentum equation, effectively enabling simulations with a lower effective Prandtl number ($P_m$).

In contrast, quasi-DNS studies of deep solar convection with $P_m$ values around $0.1$ have failed to detect SSD action \citep{2018AN....339..127K}. The work by \citep{2018AN....339..127K} demonstrated that the growth rate scaled as $\gamma \sim R_m^{1/2}$ for $P_m < 1$, which is different from the numerical simulation results generated by ideal MHD and random noise-driven turbulence mentioned above. This discrepancy is likely attributable to the influence of the large-scale motions in the deep convection dynamo model. Global and semi-global convection simulations in spherical geometry have demonstrated the existence of the SSD, but so far only for $P_m=1$. Since these simulations include rotational motion and stratification, it is difficult to distinguish whether the magnetic field fluctuations are caused by the small-scale turbulent dynamo or by the magnetic field tangling induced by the large-scale turbulent dynamo.

The anisotropic turbulent magnetic diffusion has been incorperated into a mean-field solar dynamo model to demonstrate its role in modulating the solar cycle period and the polar field distribution\citep{2014ApJ...785...49P}. Later, a hybrid LES approach nested with small-scale DNS was applied to analyze energy conversion mechanisms in the corona under extreme high-$P_m$ conditions \citep{2019ApJ...879...57B}, exploring the phenomenon of the inverse dynamo. Under conditions of a large magnetic Prandtl number, the Lorentz force does work on the fluid, driving small-scale flows and converting magnetic energy into kinetic energy, which is opposite to the energy conversion direction of the traditional small-scale turbulent dynamo. Their study found that this inverse dynamo action becomes increasingly dominant with rising $P_m$. These simulation results not only deepen our understanding of turbulent dynamo mechanisms but also provide a theoretical basis for interpreting observational data \citep{2019ApJ...879...57B}.

A recent review\citep{2023SSRv..219...36R} further emphasizes the need for larger-scale simulations incorporating more realistic physics (such as radiative transfer and partial ionization) to deeply explore the nonlinear amplification process of magnetic fields under the extreme condition of $P_m \ll$ 1. This is crucial for precisely identifying the physical processes and determining the efficiency of the small-scale turbulent dynamo.

\subsection{Research Progress in Convection-Driven Small-Scale Solar Turbulent Dynamos}
As mentioned earlier, the convection-driven small-scale turbulent dynamo was studied by solving simplified, incompressible MHD equations \citep{1999ApJ...515L..39C}. The results indicated that a substantial portion of the magnetic field in the solar quiet Sun originates from the SSD process. 

Subsequently, V\"{o}gler and Sch\"{u}ssler, for the first time, demonstrated in compressible magnetohydrodynamic (MHD) simulations that included more realistic gravitational stratification, radiative transfer, and partial ionization effects that, in an open system, as long as the magnetic Reynolds number is sufficiently large, the convection generated at the solar surface can also excite local dynamo effects and sustain magnetic field amplification \citep{2007A&A...465L..43V}. This study, through numerical simulations, proved the generation of local dynamo effects in an environment close to the real Sun. Their further research showed that the near-surface convection-driven local dynamo generates a mixed-polarity, small-scale magnetic field with a "salt-and-pepper" pattern. The closed-loop topology of these fields naturally leads to a dominance of the horizontal field component in the upper photosphere. This mechanism, operating independently of the large-scale magnetic field, is a significant source of the internetwork magnetic field in the quiet Sun \citep{2008A&A...481L...5S}. \citep{2010ApJ...714.1606P} analyzed their simulation results by using an energy balance equation and transfer functions, and the source of magnetic energy was quantified. They found that the magnetic energy primarily originates from the stretching of small-scale magnetic field lines by turbulence\citep{2010ApJ...714.1606P}. These findings strongly support the existence of a local dynamo at the solar surface. \citep{2017A&A...604A..66K} solved the non-ideal MHD equations by including the realistic equation of state, non-grey radiative transfer, and a generalized induction equation (retaining only the Biermann battery term: $\nabla pe/ene$). They found that the convective motions generate electron pressure gradients, spontaneously producing seed magnetic fields on the order of $\approx 10^{-6}$ G. Although such fields are too weak to significantly impact the saturated dynamo state, this study emphasized that fundamental physical processes indeed provide a lower limit for the quiet Sun magnetic field, independent of any external seed field. It demonstrated for the first time that solar plasma can achieve self-magnetization through the battery effect and subsequent turbulent amplification without requiring an external seed, providing a physical origin for the local dynamo\citep{2017A&A...604A..66K}.

With the rapid advancement of computational software and hardware, several international research groups have conducted high-resolution simulations of the solar small-scale turbulent dynamo that include realistic effects such as gravitational stratification, radiative transfer, and partial ionization, achieving significant progress \citep{2011A&A...533A..86C,2010ApJ...723L.149D,2015ApJ...809...84K,2010ApJ...714.1606P,2014ApJ...789..132R,2007A&A...465L..43V}. The magnetic field characteristics and structures revealed by these SSD simulations in the nonlinear saturation stage resemble those observational results in the quiet Sun regions \citep{2010A&A...513A...1D,2009ApJ...693.1728P,2008A&A...481L...5S,2013IAUS..294...95S}. For example, the numerical results from the radiative MHD code MURaM have been compared with the SUNRISE observational data, and they show a high degree of consistency in the structure, distribution, and dynamic evolution of the small-scale magnetic field \citep{2010A&A...513A...1D}. The comparison in Figure \ref{fig10} between the observed and simulated line-of-sight magnetic flux density reveals that when the simulation resolution is degraded to match that of Hinode, the simulation results closely match the observations. The decay process of magnetic fields in mixed-polarity regions on the solar surface has also been simulated by using the MURaM code \citep{2011A&A...533A..86C}. The simulation results explain several aspects of existing observations. For instance, the flux cancellation mechanism was revealed by the simulations: the granular flows push opposite polarities together, leading to reconnection above the photosphere, followed by the rapid escape of U-loops and the slow retraction of the more massive $\Omega$-loops due to greater inertia. Such results are consistent with previous observational findings \citep{2012ASPC..454...41K,1999ASPC..184..222W}.

\begin{figure}
   \centering
   \includegraphics[width=0.9\textwidth]{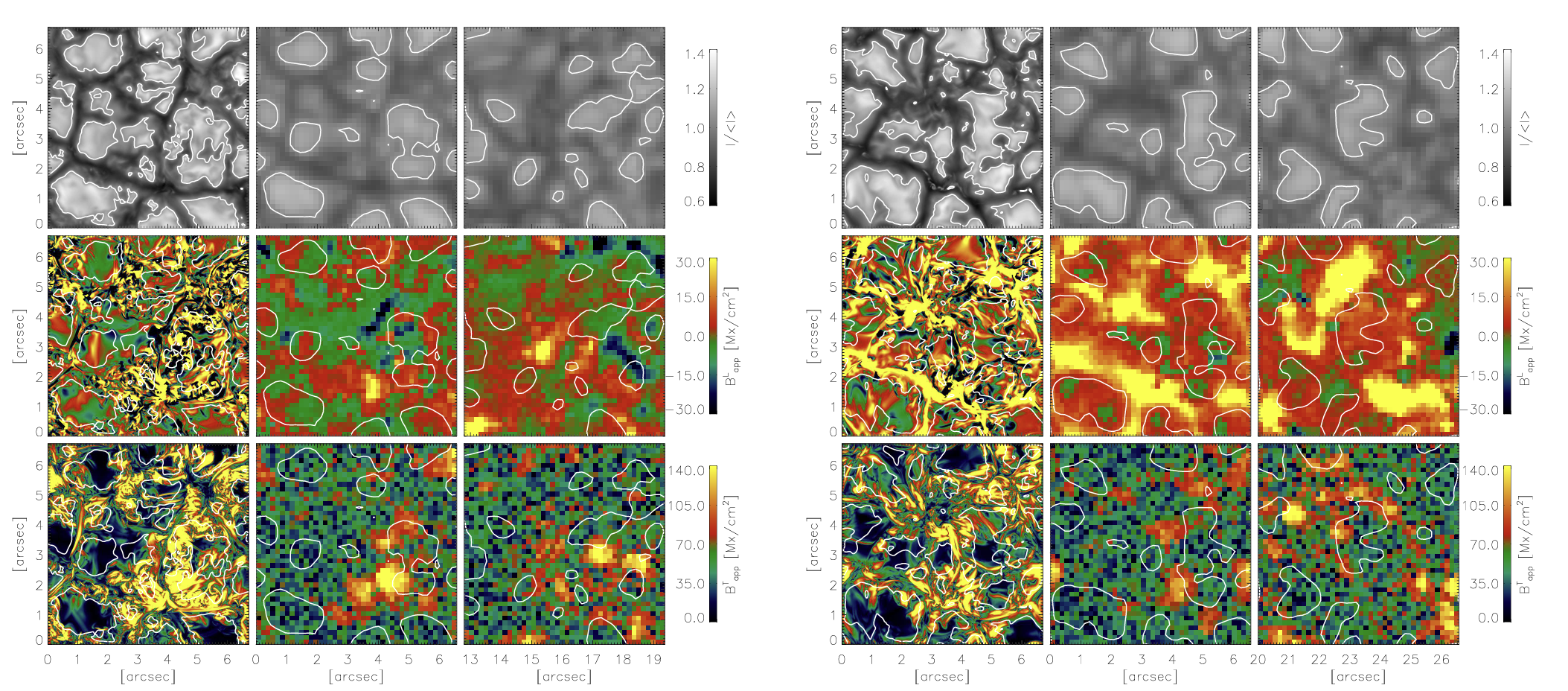}
   \caption{ Comparison of results at native resolution (left column) and Hinode resolution (middle column) with actual Hinode observations (right column), reproduced from \citet{2010A&A...513A...1D}. From top to bottom: normalized intensity, longitudinal apparent magnetic flux density, and transverse apparent magnetic flux density. The left and right maps differ only in the observed region.\newline \small Credit: S. Danilovic et al. \textit{A\&A}, Vol. 513, A1, 2010, Figs. 4 and 5. DOI: \href{http://www.aanda.org/10.1051/0004-6361/200913379}{10.1051/0004-6361/200913379}. Reproduced with permission © ESO.}
   \label{fig10}
   \end{figure}

Early simulations of small-scale turbulent dynamos in the near-surface solar layers yielded average vertical magnetic field strengths in the quiet Sun photosphere of about 20-30 G \citep{2010ApJ...723L.149D,2015ApJ...809...84K,2010ApJ...714.1606P,2007A&A...465L..43V}. These values are 2-3 times smaller than those derived from Zeeman polarimetry in the Fe I 6302 {\AA} band by the Hinode satellite, and nearly an order of magnitude smaller than results obtained from the Hanle depolarization effect in the Sr I 4607 {\AA} spectral line. Consequently, compared to observations, the simulated magnetic fields were not only too weak but also decayed too rapidly with height. The magnetic Reynolds number ($R_m$) in these simulations was significantly lower than the actual Rm in the real solar near-surface environment, which could be one factor that leads to the generated magnetic field being too weak. 

Another crucial factor relates to the bottom boundary conditions used in these simulations. Upflow-driven advection of magnetic flux was not permitted to across this boundary; flux was only allowed to leave the domain. This setup is relatively conservative, indicating that the dynamo effect can still be sustained under conditions where the local circulation is weak and magnetic energy is continuously dissipated into the deep convection zone (CZ). However, in these simulations, the influence of the deeper convection zone on the shallow-layer small-scale dynamo process is not taken into account, which does not match the actual solar environment. The deeper solar convection zone possesses a much larger $R_m$ than the photosphere, enabling the SSD to operate effectively over a broader region. \citep{2015ApJ...803...42H} simulated the SSD within a full convection zone context using an ILES approach. They found that the magnetic energy density reaches $95\%$ of the equipartition value at the base of the convection zone, indicating vigorous SSD action in the deep layers. Therefore, a portion of the magnetic field is transported to the solar surface to enhance the photospheric magnetic field and facilitate the amplification of the magnetic field in the Sun's shallow layer.

The symmetric bottom boundary conditions were then applied in his near-surface SSD simulations, which allowed horizontal magnetic fields to be advected into the domain by upflows \citep{2014ApJ...789..132R}. This modification resulted in a doubling of the photospheric magnetic field strength in the saturated stage. These improved simulations ultimately produced an average vertical field strength of about $60-80$\,G at optical depth $\tau=1$ in the photosphere \citep{2016A&A...593A..93D,2016A&A...594A.103D,2018ApJ...863..164D}, which are consistent with observational estimates based on both Zeeman and Hanle diagnostics. The Simulations of deep convective circulation \citep{2018ApJ...859..161R} show that the magnetic field upwelling from the deep layers to the photosphere (deep recirculation) undergoes significant horizontal expansion, appearing as a relatively smooth seed magnetic field at the centers of granules; whereas the magnetic field brought back to the photosphere by the mixing of local downflows/upflows (shallow recirculation) manifests as small-scale turbulent fields at the edges of granules.

As summarized in Section 2, we summarized the observational characteristics of the photospheric magnetic field in the quiet Sun region: near the solar surface, the magnetic field tends to be more isotropic, while in the mid-to-high photospheric regions, it tends to be more horizontal. The early simulations already showed a dominance of the horizontal field in the photosphere \citep{2007A&A...465L..43V}. Figure \ref{fig11} illustrates the 3D distribution of the magnetic field above the photosphere from a local dynamo simulation, one can see that the existence of a large number of magnetic loop structures leads to the magnetic field changing from vertical near the solar surface to more horizontal at greater heights \citep{2015ApJ...809...84K}. Then, a simulation with the larger domain and extending from $1.5$ Mm above the surface to the deeper convection zone was performed, it was found that the ratio of average horizontal to vertical magnetic field strength could reach 5 during the kinematic growth phase, decreasing to 2.5 in the nonlinear saturation stage \citep{2014ApJ...789..132R}. In the kinematic phase, both horizontal and vertical field components decay with height, but the vertical component decays faster. The ratio of horizontal to vertical field peaks at a height of about 450 km above the surface. The dominance of the horizontal field in the middle/upper photosphere is currently explained by two possible mechanisms: (1) Since the small-scale magnetic field generated by the SSD has mixed polarity, the maximum height of magnetic loops connecting nearby opposite-polarity patches is comparable to their foot point separation \citep{2008A&A...481L...5S}; (2) The magnetic flux expulsion mechanism, where granular convective overturning flow concentrates horizontal field in the mid-photosphere \citep{2008ApJ...680L..85S,2012ASPC..456....3S}. However, based on simulation results from \citep{2014ApJ...789..132R}, the greatest velocity field anisotropy occurs 200 km below the peak of the magnetic field anisotropy, and the height of the magnetic field anisotropy peak exactly coincides with the minimum of the convective root-mean-square velocity. Combined with the significant peak in horizontal field strength, this suggests that the turbulent pumping, specifically its diamagnetic component, might be the mechanism responsible for expelling horizontal field from the photosphere and injecting it into the low chromosphere, where it accumulates in the region of weakest turbulence. This interpretation needs further validation from future observations and simulations.

\begin{figure}
   \centering
   \includegraphics[width=0.6\textwidth]{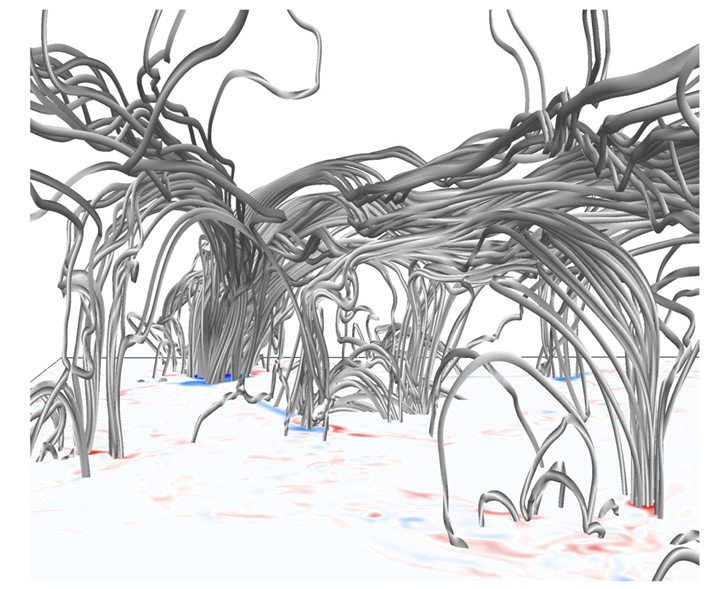}
   \caption{ Magnetic field distribution above the photospheric surface from a solar local dynamo simulation, reproduced from \citet{2015ApJ...809...84K}. The thick gray lines represent magnetic field lines. The red and blue colors at the bottom represent the vertical magnetic field strength at the solar photospheric surface, with red indicating positive polarity and blue indicating negative polarity. The magnetic field strength ranges from $-800$\,G to $300$\,G.\newline \small Source: I. N. Kitiashvili et al., ``Realistic Modeling of Local Dynamo Processes on the Sun,'' \textit{The Astrophysical Journal}, Vol. 809, No. 1, Article 84, pp. 1–12, 2015 (Figure 17). DOI: \href{https://doi.org/10.1088/0004-637X/809/1/84}{10.1088/0004-637X/809/1/84}. © AAS. Reproduced with permission.}
   \label{fig11}
   \end{figure}

The simulation results from \cite{2014ApJ...789..132R} revealed that in the nonlinear saturation stage, Lorentz force feedback suppresses kinetic energy on scales below approximately 100 km. Kilogauss-strength magnetic fields constitute a certain fraction, with these small-scale, strong fields typically appearing in locations of magnetic flux concentration within intergranular lanes. They cover approximately 0.5\% of the solar surface area and manifest as MBPs in continuum images. Figure 12 displays results from two solar turbulent dynamo simulations at different spatial scales\citep{2014ApJ...789..132R}. Apart from the simulation box size and the depth of the convection zone included, other parameters such as grid spacing and boundary conditions were identical between these runs. The figure shows that MBPs only appear within the intergranular lanes of the larger-scale simulation. Furthermore, the larger-scale simulation clearly exhibits a network magnetic field structure, which is absent in the smaller-scale simulation shown in the bottom-left panel. The author argues that accounting for the effects of deep recirculation is key to reproducing the observed photospheric network structure in simulations. They conclude that if the advection of magnetic field into the computational domain by upflows at the bottom boundary is not considered, a network structure cannot form, regardless of how deep the simulated domain extends.

\begin{figure}
   \centering
   \includegraphics[width=0.9\textwidth]{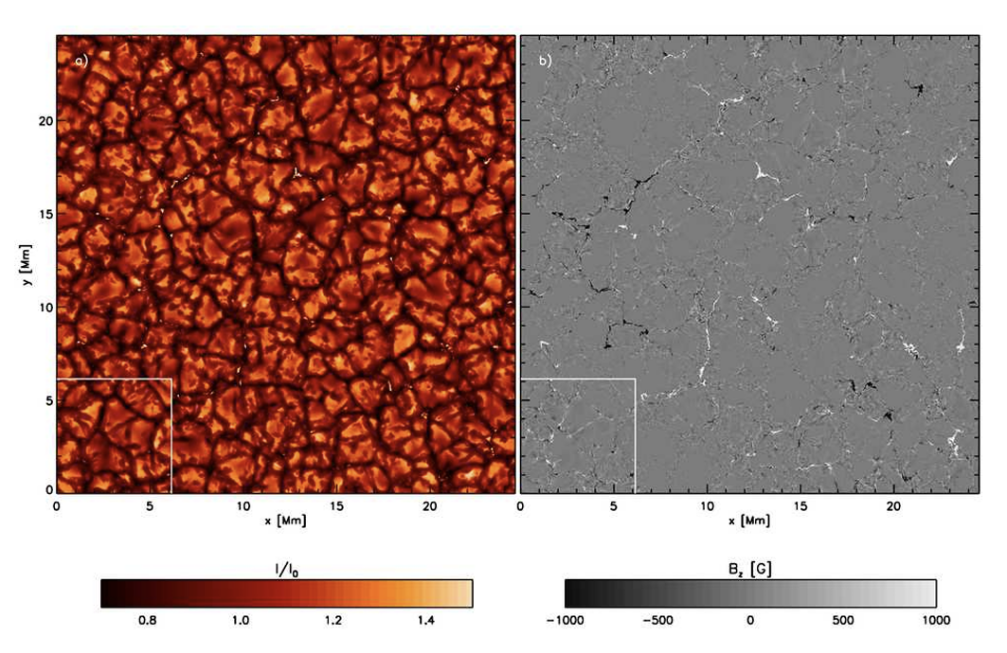}
   \caption{The simulation results reproduced from \citet{2014ApJ...789..132R}, which include two cases with different scale depths but all other parameters identical, the small-scale simulation result is displayed within a white box in the lower-left corner of the large-scale simulation result. (a) shows the intensity distribution. (b) shows the vertical magnetic field component (Bz) at optical depth $\tau=1$ \citep{2014ApJ...789..132R}. \newline \small Source: M. Rempel, ``Numerical Simulations of Quiet Sun Magnetism: On the Contribution from a Small-Scale Dynamo,'' \textit{The Astrophysical Journal}, Vol. 789, No. 2, Article 132, pp. 1–22, 2014 (Figure 8). DOI: \href{https://doi.org/10.1088/0004-637X/789/2/132}{10.1088/0004-637X/789/2/132}. © AAS. Reproduced with permission.}
   \label{fig12}
   \end{figure}

Although the small-scale magnetic fields in the quiet Sun are mostly too weak to affect the radiative properties of the photosphere, a small number of kilogauss (kG) magnetic flux concentration regions (about $0.5\%$) exist in the photosphere, which, by enhancing radiative losses, make the radiative loss rate comparable to the magnetic flux concentration effect in the solar network region (i.e., the region where magnetic flux is significantly unbalanced in the photosphere). \citep{2020ApJ...894..140R} calculated the dependence of the solar radiative energy flux on the average vertical magnetic field strength using a series of small-scale dynamo simulation results with different magnetic field strengths. It was found that at optical depth $\tau=1$ in the photosphere, for every 10 Gauss change in the mean vertical field strength, the Total Solar Irradiance (TSI) changes by approximately 0.14\% (for the pure small-scale dynamo case) to 0.173\% (for the small-scale dynamo + network field case). Even these relatively small magnetic variations could induce TSI changes comparable to those observed over the solar cycle. This indicates their global impact on convective energy distribution and highlights the potential importance of small-scale magnetic field variations in long-term solar irradiance changes.

\section{Summary and Future Perspectives}

\subsection{Summary}

The small-scale (or local) dynamo is regarded as the central mechanism explaining the origin of the persistent, diffuse weak magnetic field in the quiet Sun. Significant progress over the past three decades in observations, theory, and numerical simulations has profoundly reshaped our understanding of solar magnetism. This review has systematically outlined the research framework in this field, with the core conclusion being that a small-scale turbulent dynamo process, independent of the solar cycle and driven by turbulent convection, operates within the quiet photosphere and convection zone. The magnetic fields generated by this process possess distinct statistical characteristics (e.g.high intermittency, multi-scale distribution) and, during solar minimum, contribute the majority of the Sun's magnetic flux \citep{2011ApJ...731...37J}. Numerical simulations have successfully reproduced this turbulent dynamo process, demonstrating that turbulent stretching, folding, and twisting of magnetic fields can effectively amplify and sustain magnetism even in the solar plasma with an extremely low magnetic Prandtl number ($P_m \ll 1$) \citep{2011ApJ...741...92B,2011ApJ...727L..50K}. Observationally, high-resolution space-based and ground-based facilities—such as Hinode/SOT, SDO/HMI, SUNRISE, SST, GREGOR, GST, and DKIST—have provided direct evidence supporting the existence of the small-scale dynamo, including the ubiquitous horizontal fields in the quiet Sun, frequently emerging small-scale magnetic loops, and the formation and evolution of MBPs \citep{2017SSRv..210..275B,2023ApJ...955L..36C,2008ApJ...672.1237L,2010ApJ...714L..94M,2007ApJ...670L..61O,2010ApJ...723L.127S}. Theoretical studies indicate that the magnetic fields generated by the small-scale dynamo are not only the source of most magnetic energy in the quiet Sun but also profoundly influence energy transport and dynamic processes in the photosphere and chromosphere \citep{2020ApJ...894..140R}. Furthermore, by interacting with the active region fields produced by the global solar dynamo, they collectively shape the complex panorama of solar magnetism \citep{2019ApJ...871...16J}.

\subsection{Future Perspectives}

Despite remarkable achievements, research on the small-scale solar dynamo faces a series of key challenges and cutting-edge directions that urgently require further exploration:

 \textbf{1. Enhancing Observational Precision and Multi-instrument Synergy:} Next-generation facilities, exemplified by the Daniel K. Inouye Solar Telescope (DKIST) are probing the quiet Sun magnetic field with unprecedented spatial resolution (~20 km) and sensitivity. DKIST has begun to reveal the fine-scale topology of the quiet Sun magnetic field, with its core potential for future growth lying in multi-wavelength, simultaneous polarimetric capabilities \citep{2025ApJ...979..139L}. Future observations must overcome single-height limitations to achieve synchronous, high-resolution measurements of magnetic, velocity, and thermodynamic parameters across multiple atmospheric layers-from the photosphere and chromosphere to the transition region \citep{2024ApJ...964...27R}. This requires advanced real-time adaptive optics (AO) image reconstruction techniques, and multi-instrument observational strategies to precisely track energy dissipation processes \citep{2024ApJ...964...27R}. Precise determination of the behavior of the magnetic energy spectrum at scales smaller than $10$\,km will be a key observational test to distinguish between a pure small-scale dynamo and the fragmentation of the global dynamo \citep{2014A&A...572A..98A}. Utilizing synchronized observations from DKIST and other instruments to precisely quantify the transition of magnetic energy dominance from the photosphere to the chromosphere \citep{2014A&A...572A..98A} and exploring how the rapid migration of anti-correlated magnetic elements \citep{2012ApJ...745...39J} affects the cross-scale coupling model, are also vital future directions for observational studies of the solar turbulent dynamo. China's upcoming 2-meter Ring Solar Telescope (2mRST), 2.5-meter WeHoST, and the future 8-meter Chinese Giant Solar Telescope (CGST) and the proposed Solar Close Observations and Proximity Exploration mission (SCOPE) \citep{2025AstTI...2..148L} are expected to provide substantial observational data for small-scale dynamo research. Furthermore, careful instrumental noise filtering should be a standard procedure in future high-resolution observations of quiet-Sun magnetism, as it is essential for unambiguously identifying SSD signatures \citep{2012A&A...541A..17S}.
 
 \textbf{2. Advancing Theoretical Models and Refining Numerical Simulations:} Current numerical simulations still have substantial room for improvement in incorporating realistic physical processes (e.g., radiative transfer, partial ionization, non-ideal MHD effects) and in expanding the parameter space (e.g., lower $P_m$, larger magnetic Reynolds number $R_m$). The saturation mechanism and efficiency of the turbulent dynamo at low magnetic Prandtl number ($P_m$) remain core theoretical challenges\citep{2007PhRvL..98t8501I,2007NJPh....9..300S}. Since the real value of $P_m$ in the solar photosphere and convection zone is extremely low ($10^{-6} \sim 10^{-4}$) \citep{2020RvMP...92d1001S}, numerical simulations require extremely high spatial resolution to reach such low $P_m$ values. As mentioned earlier, a small-scale dynamo at $P_m \approx 0.003$ \citep{2023NatAs...7..662W} was successfully excited, which represents the lowest $P_m$ at which an SSD has been achieved in numerical simulations to date. However, those simulations did not include physical processes such as gravitational stratification or radiative transfer, and therefore lacked realistic solar convection. Furthermore, those studies only investigated the kinematic stage of the SSD. Currently, simulations that include realistic gravitational stratification, radiative transfer, and convection-driven turbulence have only reached a minimum $P_m$ value close to $0.1$ for exciting the small-scale turbulent dynamo. Therefore, future efforts must focus on developing more efficient algorithms and leveraging larger-scale computations to explore the nonlinear amplification of magnetic fields, their saturation levels, and their feedback on convective structures under the extreme condition of $P_m \ll 1$ \citep{2023SSRv..219...36R}. At the same time, it is necessary to more self-consistently couple radiative magnetohydrodynamic (rMHD) processes to deeply investigate how magnetic fields affect convective energy transport and radiation intensity variations \citep{2020ApJ...894..140R}, as well as to study the effects of non-ideal MHD phenomena such as magnetic diffusion caused by neutral particles, ambipolar diffusion, and the Hall effect on the structure and evolution of small-scale magnetic fields \citep{2024ApJ...964...27R,2025A&A...697A..29K}.
 
The internetwork magnetic field in the quiet Sun is often considered weakly linked to the global dynamo process. However, the observed solar cycle variation of the network field indicates that, besides the small-scale turbulent dynamo, the quiet Sun network field is also related to the global dynamo process. Our understanding of the strength of the large-scale turbulent dynamo and its evolution throughout the solar cycle remains limited. The interaction between the small-scale turbulent dynamo and the large-scale global dynamo is a challenging interdisciplinary frontier, requiring the construction of coupled models spanning spatial and temporal scales \citep{2024IAUS..365..261A}. Such models are essential to elucidate how small-scale fields in the quiet Sun are transported by large-scale organized flows, such as differential rotation or meridional circulation \citep{2025A&A...696A.143R} and how the background field generated by the global dynamo \citep{2024ApJ...961L..46C} can suppress local dynamo efficiency \citep{2005PhR...417....1B,2019ApJ...871...16J}. Understanding this cross-scale coupling is crucial for building a unified picture of solar magnetic field evolution and energy cycling. Currently, a limited number of global and semi-global large-scale simulations can simultaneously excite both the large-scale (LSD) and small-scale (SSD) dynamos\citep{2016Sci...351.1427H,2017A&A...599A...4K}, but results differ significantly between models. Employing more comprehensive and accurate simulations to deeply investigate the connection between LSD and SSD is a crucial future direction for solar dynamo research.

\textbf{3. Extending to Stellar Applications and the Sun-Stellar Connection: }The Sun, as the only star whose surface magnetic field can be resolved in detail, serves as a fundamental laboratory for studying stellar magnetic activity. The success of small-scale dynamo theory provides a foundation for understanding the origin of surface magnetism in stars with convective envelopes, particularly for those with lower activity and fully convection \citep{2023SSRv..219...36R}. In the future, through higher-resolution, cross-scale simulations and stellar analogies, the SSD needs to be advanced from observational signatures to a predictable and quantifiable physical mechanism, and its true role in the magnetic field evolution of the Sun and stars needs to be clarified \citep{2024IAUS..365..261A}. Research needs to extrapolate solar-validated models to different stellar parameter spaces (mass, convective envelope depth, rotation rate) to predict surface magnetic field strength distributions, energy spectra, and their implications for stellar flares, coronal heating, and stellar winds. Detailed solar observations, especially from high-resolution facilities like DKIST, and future Chinese telescopes like 2-meter Ring Solar Telescope (2mRST), the 2.5-meter WeHoST and 8-meter CGST, will provide critical calibration and constraints for these models. Conversely, observations of stellar magnetic activity (such as studies of super-quiet stars and magnetic activity "minima") can provide important evidence and predictions for understanding the Sun's position among solar-type stars and exploring the magnetic state of the Sun during potential future super-quiet periods.

Research on the solar small-scale dynamo has progressed from theoretical hypothesis to observational confirmation and numerical reproduction, marking its significant and profound role in exploring the origin of the Sun's magnetic field. With the commissioning of more advanced facilities, continuous leaps in computational capabilities, and deeper interdisciplinary research, this field promises substantial development in the coming decades. Breaking through the understanding of the physical nature of the turbulent dynamo at low $P_m$, achieving self-consistent coupling of magnetic fields across scales, precisely quantifying turbulent transport effects, and extending knowledge from the Sun to the stellar world represent the forefront and key directions for future small-scale turbulent dynamo research. Breakthroughs in these areas will not only help us further unravel the mysteries of the quiet Sun's magnetic field but also provide a crucial physical foundation for understanding stellar magnetic activity, the origins of space weather, and the universal principles of plasma turbulent dynamos.

 \normalem
\begin{acknowledgements}
This research is supported by the Strategic Priority Research Program of the Chinese Academy of Sciences (Grant No.XDB0560000); the National Natural Science Foundation of China  (Grants No.12373060 and No.12473053 and 12473054); the National Key R\&D Program of China (Grant No.2022YFF0503804(2022YFF0503800) and No.2022YFF0503003 (2022YFF0503000)); the outstanding member of the Youth Innovation Promotion Association CAS (No.Y2021024); the Basic Research of Yunnan Province in China  (Grant No.202401AS070044); the Yunling Talent Project for the Youth; the Yunling Scholar Project of the Yunnan Province and the Yunnan Province Scientist Workshop of Solar Physics; Yunnan Key Laboratory of Solar Physics and Space Science (No. 202205AG070009). 
\end{acknowledgements}

\bibliographystyle{raa}
\bibliography{bibtex}

\begin{thebibliography}{183}
\providecommand\natexlab[1]{#1}
\providecommand\JournalTitle[1]{#1}

\bibitem[{Abramenko}(2024)]{2024IAUS..365..261A}
{Abramenko}, V.~I. 2024, in IAU Symposium, Vol. 365, Dynamics of Solar and
  Stellar Convection Zones and Atmospheres, ed. A.~V. {Getling} \& L.~L.
  {Kitchatinov}, 261

\bibitem[{Amari} {et~al.}(2015)]{2015Natur.522..188A}
{Amari}, T., {Luciani}, J.-F., \& {Aly}, J.-J. 2015, \nat, 522, 188

\bibitem[{Asensio Ramos}(2009)]{2009ApJ...701.1032A}
{Asensio Ramos}, A. 2009, \apj, 701, 1032

\bibitem[{Asensio Ramos} \& {Mart{\'\i}nez
  Gonz{\'a}lez}(2014)]{2014A&A...572A..98A}
{Asensio Ramos}, A., \& {Mart{\'\i}nez Gonz{\'a}lez}, M.~J. 2014, \aap, 572,
  A98

\bibitem[{Batchelor}(1950)]{1950QJRMS..76..133B}
{Batchelor}, G.~K. 1950, Quarterly Journal of the Royal Meteorological Society,
  76, 133

\bibitem[{Bellot Rubio} \& {Orozco Su{\'a}rez}(2019)]{2019LRSP...16....1B}
{Bellot Rubio}, L., \& {Orozco Su{\'a}rez}, D. 2019, Living Reviews in Solar
  Physics, 16, 1

\bibitem[{Berghmans} {et~al.}(2021)]{2021A&A...656L...4B}
{Berghmans}, D., {Auch{\`e}re}, F., {Long}, D.~M., {et~al.} 2021, \aap, 656, L4

\bibitem[{Bonet} {et~al.}(2012)]{2012A&A...539A...6B}
{Bonet}, J.~A., {Cabello}, I., \& {S{\'a}nchez Almeida}, J. 2012, \aap, 539, A6

\bibitem[{Borrero} {et~al.}(2017)]{2017SSRv..210..275B}
{Borrero}, J.~M., {Jafarzadeh}, S., {Sch{\"u}ssler}, M., \& {Solanki}, S.~K.
  2017, \ssr, 210, 275

\bibitem[{Borrero} \& {Kobel}(2011)]{2011A&A...527A..29B}
{Borrero}, J.~M., \& {Kobel}, P. 2011, \aap, 527, A29

\bibitem[{Borrero} \& {Kobel}(2013)]{2013A&A...550A..98B}
{Borrero}, J.~M., \& {Kobel}, P. 2013, \aap, 550, A98

\bibitem[{Borrero} {et~al.}(2010)]{2010ApJ...723L.144B}
{Borrero}, J.~M., {Mart{\'\i}nez-Pillet}, V., {Schlichenmaier}, R., {et~al.}
  2010, \apjl, 723, L144

\bibitem[{Brandenburg}(2011)]{2011ApJ...741...92B}
{Brandenburg}, A. 2011, \apj, 741, 92

\bibitem[{Brandenburg} \& {Rempel}(2019)]{2019ApJ...879...57B}
{Brandenburg}, A., \& {Rempel}, M. 2019, \apj, 879, 57

\bibitem[{Brandenburg} {et~al.}(2012)]{2012SSRv..169..123B}
{Brandenburg}, A., {Sokoloff}, D., \& {Subramanian}, K. 2012, \ssr, 169, 123

\bibitem[{Brandenburg} \& {Subramanian}(2005)]{2005PhR...417....1B}
{Brandenburg}, A., \& {Subramanian}, K. 2005, \physrep, 417, 1

\bibitem[{Buehler} {et~al.}(2013)]{2013A&A...555A..33B}
{Buehler}, D., {Lagg}, A., \& {Solanki}, S.~K. 2013, \aap, 555, A33

\bibitem[{Cameron} {et~al.}(2011)]{2011A&A...533A..86C}
{Cameron}, R., {V{\"o}gler}, A., \& {Sch{\"u}ssler}, M. 2011, \aap, 533, A86

\bibitem[{Campbell} {et~al.}(2023)]{2023ApJ...955L..36C}
{Campbell}, R.~J., {Keys}, P.~H., {Mathioudakis}, M., {et~al.} 2023, \apjl,
  955, L36

\bibitem[{Cao} {et~al.}(2010)]{2010AN....331..636C}
{Cao}, W., {Gorceix}, N., {Coulter}, R., {et~al.} 2010, Astronomische
  Nachrichten, 331, 636

\bibitem[{Cattaneo}(1999)]{1999ApJ...515L..39C}
{Cattaneo}, F. 1999, \apjl, 515, L39

\bibitem[{Centeno} {et~al.}(2007)]{2007ApJ...666L.137C}
{Centeno}, R., {Socas-Navarro}, H., {Lites}, B., {et~al.} 2007, \apjl, 666,
  L137

\bibitem[{Charbonneau}(2020)]{2020LRSP...17....4C}
{Charbonneau}, P. 2020, Living Reviews in Solar Physics, 17, 4

\bibitem[{Chen} {et~al.}(2021)]{2021A&A...656L...7C}
{Chen}, Y., {Przybylski}, D., {Peter}, H., {et~al.} 2021, \aap, 656, L7

\bibitem[{Chernyshov} {et~al.}(2008)]{2008ApJ...686.1137C}
{Chernyshov}, A.~A., {Karelsky}, K.~V., \& {Petrosyan}, A.~S. 2008, \apj, 686,
  1137

\bibitem[{Chernyshov} {et~al.}(2009)]{2009essu.confE..75C}
{Chernyshov}, D., {Cheng}, K.~S., {Dogiel}, V., {Ko}, C.~M., \& {Ip}, W.~H.
  2009, in The Extreme Sky: Sampling the Universe above 10 keV, 75

\bibitem[{Chernyshov} {et~al.}(2014)]{2014arXiv1412.2740C}
{Chernyshov}, D.~O., {Leung}, G. C.~K., {Cheng}, K.~S., {Dogiel}, V.~A., \&
  {Tatischeff}, V. 2014, arXiv e-prints, arXiv:1412.2740

\bibitem[{Cheung} \& {Isobe}(2014)]{2014LRSP...11....3C}
{Cheung}, M. C.~M., \& {Isobe}, H. 2014, Living Reviews in Solar Physics, 11, 3

\bibitem[{Chitta} {et~al.}(2023)]{2023Sci...381..867C}
{Chitta}, L.~P., {Zhukov}, A.~N., {Berghmans}, D., {et~al.} 2023, Science, 381,
  867

\bibitem[{Choudhuri}(2013)]{2013IAUS..294...37C}
{Choudhuri}, A.~R. 2013, in IAU Symposium, Vol. 294, Solar and Astrophysical
  Dynamos and Magnetic Activity, ed. A.~G. {Kosovichev}, E.~{de Gouveia Dal
  Pino}, \& Y.~{Yan}, 37

\bibitem[{Cliver} {et~al.}(2024)]{2024ApJ...961L..46C}
{Cliver}, E.~W., {White}, S.~M., \& {Richardson}, I.~G. 2024, \apjl, 961, L46

\bibitem[{Danilovic} {et~al.}(2016{\natexlab{a}})]{2016A&A...594A.103D}
{Danilovic}, S., {Rempel}, M., {van Noort}, M., \& {Cameron}, R.
  2016{\natexlab{a}}, \aap, 594, A103

\bibitem[{Danilovic} {et~al.}(2010{\natexlab{a}})]{2010A&A...513A...1D}
{Danilovic}, S., {Sch{\"u}ssler}, M., \& {Solanki}, S.~K. 2010{\natexlab{a}},
  \aap, 513, A1

\bibitem[{Danilovic} {et~al.}(2016{\natexlab{b}})]{2016A&A...593A..93D}
{Danilovic}, S., {van Noort}, M., \& {Rempel}, M. 2016{\natexlab{b}}, \aap,
  593, A93

\bibitem[{Danilovic} {et~al.}(2010{\natexlab{b}})]{2010ApJ...723L.149D}
{Danilovic}, S., {Beeck}, B., {Pietarila}, A., {et~al.} 2010{\natexlab{b}},
  \apjl, 723, L149

\bibitem[{de Wijn} {et~al.}(2008)]{2008ApJ...684.1469D}
{de Wijn}, A.~G., {Lites}, B.~W., {Berger}, T.~E., {et~al.} 2008, \apj, 684,
  1469

\bibitem[{del Pino Alem{\'a}n} {et~al.}(2018)]{2018ApJ...863..164D}
{del Pino Alem{\'a}n}, T., {Trujillo Bueno}, J., {{\v{S}}t{\v{e}}p{\'a}n}, J.,
  \& {Shchukina}, N. 2018, \apj, 863, 164

\bibitem[{Dikpati}(2011)]{2011ApJ...733...90D}
{Dikpati}, M. 2011, \apj, 733, 90

\bibitem[{Erd{\'e}lyi} \& {Fedun}(2007)]{2007Sci...318.1572E}
{Erd{\'e}lyi}, R., \& {Fedun}, V. 2007, Science, 318, 1572

\bibitem[{Fan}(2008)]{2008ApJ...676..680F}
{Fan}, Y. 2008, \apj, 676, 680

\bibitem[{Fan}(2009)]{2009LRSP....6....4F}
{Fan}, Y. 2009, Living Reviews in Solar Physics, 6, 4

\bibitem[{Fan} \& {Fang}(2014)]{2014ApJ...789...35F}
{Fan}, Y., \& {Fang}, F. 2014, \apj, 789, 35

\bibitem[{Golub} {et~al.}(1979)]{1979ApJ...229L.145G}
{Golub}, L., {Davis}, J.~M., \& {Krieger}, A.~S. 1979, \apjl, 229, L145

\bibitem[{Go{\v{s}}i{\'c}} {et~al.}(2016)]{2016ApJ...820...35G}
{Go{\v{s}}i{\'c}}, M., {Bellot Rubio}, L.~R., {del Toro Iniesta}, J.~C.,
  {Orozco Su{\'a}rez}, D., \& {Katsukawa}, Y. 2016, \apj, 820, 35

\bibitem[{Go{\v{s}}i{\'c}} {et~al.}(2014)]{2014ApJ...797...49G}
{Go{\v{s}}i{\'c}}, M., {Bellot Rubio}, L.~R., {Orozco Su{\'a}rez}, D.,
  {Katsukawa}, Y., \& {del Toro Iniesta}, J.~C. 2014, \apj, 797, 49

\bibitem[{Grete} {et~al.}(2017)]{2017PhRvE..95c3206G}
{Grete}, P., {Vlaykov}, D.~G., {Schmidt}, W., \& {Schleicher}, D. R.~G. 2017,
  \pre, 95, 033206

\bibitem[{Grinstein} {et~al.}(2011)]{2011PhFl...23c4106G}
{Grinstein}, F.~F., {Gowardhan}, A.~A., \& {Wachtor}, A.~J. 2011, Physics of
  Fluids, 23, 034106

\bibitem[{Hagenaar} {et~al.}(2003)]{2003ApJ...584.1107H}
{Hagenaar}, H.~J., {Schrijver}, C.~J., \& {Title}, A.~M. 2003, ApJ, 584, 1107

\bibitem[{Harvey} {et~al.}(2007)]{2007ApJ...659L.177H}
{Harvey}, J.~W., {Branston}, D., {Henney}, C.~J., {Keller}, C.~U., \& {SOLIS
  and GONG Teams}. 2007, \apjl, 659, L177

\bibitem[{Harvey} \& {Harvey}(1973)]{1973SoPh...28...61H}
{Harvey}, K., \& {Harvey}, J. 1973, \solphys, 28, 61

\bibitem[{Hazra} {et~al.}(2023)]{2023SSRv..219...39H}
{Hazra}, G., {Nandy}, D., {Kitchatinov}, L., \& {Choudhuri}, A.~R. 2023, \ssr,
  219, 39

\bibitem[{Hotta} {et~al.}(2022)]{2022ApJ...933..199H}
{Hotta}, H., {Kusano}, K., \& {Shimada}, R. 2022, \apj, 933, 199

\bibitem[{Hotta} {et~al.}(2014)]{2014ApJ...786...24H}
{Hotta}, H., {Rempel}, M., \& {Yokoyama}, T. 2014, \apj, 786, 24

\bibitem[{Hotta} {et~al.}(2015)]{2015ApJ...803...42H}
{Hotta}, H., {Rempel}, M., \& {Yokoyama}, T. 2015, \apj, 803, 42

\bibitem[{Hotta} {et~al.}(2016)]{2016Sci...351.1427H}
{Hotta}, H., {Rempel}, M., \& {Yokoyama}, T. 2016, Science, 351, 1427

\bibitem[{Hou} {et~al.}(2024)]{2024NatAs...8.1246H}
{Hou}, C., {He}, J., {Duan}, D., {et~al.} 2024, Nature Astronomy, 8, 1246

\bibitem[{Huang} {et~al.}(2014)]{2014ApJ...797...88H}
{Huang}, Z., {Madjarska}, M.~S., {Xia}, L., {et~al.} 2014, \apj, 797, 88

\bibitem[{Iida} {et~al.}(2012)]{2012ApJ...752..149I}
{Iida}, Y., {Hagenaar}, H.~J., \& {Yokoyama}, T. 2012, \apj, 752, 149

\bibitem[{Iida} {et~al.}(2015)]{2015ApJ...814..134I}
{Iida}, Y., {Hagenaar}, H.~J., \& {Yokoyama}, T. 2015, \apj, 814, 134

\bibitem[{Ishikawa} \& {Tsuneta}(2009)]{2009A&A...495..607I}
{Ishikawa}, R., \& {Tsuneta}, S. 2009, \aap, 495, 607

\bibitem[{Ishikawa} {et~al.}(2010)]{2010ApJ...713.1310I}
{Ishikawa}, R., {Tsuneta}, S., \& {Jur{\v{c}}{\'a}k}, J. 2010, \apj, 713, 1310

\bibitem[{Iskakov} {et~al.}(2007)]{2007PhRvL..98t8501I}
{Iskakov}, A.~B., {Schekochihin}, A.~A., {Cowley}, S.~C., {McWilliams}, J.~C.,
  \& {Proctor}, M.~R.~E. 2007, \prl, 98, 208501

\bibitem[{Jiang} {et~al.}(2014)]{2014SSRv..186..491J}
{Jiang}, J., {Hathaway}, D.~H., {Cameron}, R.~H., {et~al.} 2014, \ssr, 186, 491

\bibitem[{Jiang} {et~al.}(2019)]{2019ApJ...871...16J}
{Jiang}, J., {Song}, Q., {Wang}, J.-X., \& {Baranyi}, T. 2019, \apj, 871, 16

\bibitem[{Jin} \& {Wang}(2012)]{2012ApJ...745...39J}
{Jin}, C.~L., \& {Wang}, J.~X. 2012, \apj, 745, 39

\bibitem[{Jin} \& {Wang}(2015{\natexlab{a}})]{2015ApJ...807...70J}
{Jin}, C.~L., \& {Wang}, J.~X. 2015{\natexlab{a}}, \apj, 807, 70

\bibitem[{Jin} \& {Wang}(2019)]{2019RAA....19...69J}
{Jin}, C.-L., \& {Wang}, J.-X. 2019, Research in Astronomy and Astrophysics,
  19, 069

\bibitem[{Jin} {et~al.}(2011)]{2011ApJ...731...37J}
{Jin}, C.~L., {Wang}, J.~X., {Song}, Q., \& {Zhao}, H. 2011, \apj, 731, 37

\bibitem[{Jin} \& {Wang}(2015{\natexlab{b}})]{2015ApJ...806..174J}
{Jin}, C., \& {Wang}, J. 2015{\natexlab{b}}, \apj, 806, 174

\bibitem[{Jin} {et~al.}(2009)]{2009ApJ...690..279J}
{Jin}, C., {Wang}, J., \& {Zhao}, M. 2009, \apj, 690, 279

\bibitem[{Jin} {et~al.}(2026)]{2026ApJS..283...33J}
{Jin}, C., {Zhou}, G., {Hu}, J., \& {Wang}, J. 2026, ApJS, 283, 33

\bibitem[{K{\"a}pyl{\"a}} {et~al.}(2018)]{2018AN....339..127K}
{K{\"a}pyl{\"a}}, P.~J., {K{\"a}pyl{\"a}}, M.~J., \& {Brandenburg}, A. 2018,
  Astronomische Nachrichten, 339, 127

\bibitem[{K{\"a}pyl{\"a}} {et~al.}(2017)]{2017A&A...599A...4K}
{K{\"a}pyl{\"a}}, P.~J., {K{\"a}pyl{\"a}}, M.~J., {Olspert}, N., {Warnecke},
  J., \& {Brandenburg}, A. 2017, \aap, 599, A4

\bibitem[{Kazantsev}(1968)]{1968JETP...26.1031K}
{Kazantsev}, A.~P. 1968, Soviet Journal of Experimental and Theoretical
  Physics, 26, 1031

\bibitem[{Khomenko} {et~al.}(2003)]{2003A&A...408.1115K}
{Khomenko}, E.~V., {Collados}, M., {Solanki}, S.~K., {Lagg}, A., \& {Trujillo
  Bueno}, J. 2003, \aap, 408, 1115

\bibitem[{Khomenko} {et~al.}(2017)]{2017A&A...604A..66K}
{Khomenko}, E., {Vitas}, N., {Collados}, M., \& {de Vicente}, A. 2017, \aap,
  604, A66

\bibitem[{Khomenko} {et~al.}(2025)]{2025A&A...697A..29K}
{Khomenko}, E., {Vitas}, N., {Collados}, M., \& {Modestov}, M. 2025, \aap, 697,
  A29

\bibitem[{Kitchatinov}(2026)]{2026arXiv260401718K}
{Kitchatinov}, L. 2026, arXiv e-prints, arXiv:2604.01718

\bibitem[{Kitiashvili} {et~al.}(2011)]{2011ApJ...727L..50K}
{Kitiashvili}, I.~N., {Kosovichev}, A.~G., {Mansour}, N.~N., \& {Wray}, A.~A.
  2011, \apjl, 727, L50

\bibitem[{Kitiashvili} {et~al.}(2015)]{2015ApJ...809...84K}
{Kitiashvili}, I.~N., {Kosovichev}, A.~G., {Mansour}, N.~N., \& {Wray}, A.~A.
  2015, \apj, 809, 84

\bibitem[{Kleeorin} \& {Rogachevskii}(2012)]{2012PhyS...86a8404K}
{Kleeorin}, N., \& {Rogachevskii}, I. 2012, \physscr, 86, 018404

\bibitem[{Kobel} {et~al.}(2011)]{2011A&A...531A.112K}
{Kobel}, P., {Solanki}, S.~K., \& {Borrero}, J.~M. 2011, \aap, 531, A112

\bibitem[{Kobel} {et~al.}(2012)]{2012A&A...542A..96K}
{Kobel}, P., {Solanki}, S.~K., \& {Borrero}, J.~M. 2012, \aap, 542, A96

\bibitem[{Kubo} {et~al.}(2012)]{2012ASPC..454...41K}
{Kubo}, M., {Low}, B.~C., \& {Lites}, B.~W. 2012, in Astronomical Society of
  the Pacific Conference Series, Vol. 454, Hinode-3: The 3rd Hinode Science
  Meeting, ed. T.~{Sekii}, T.~{Watanabe}, \& T.~{Sakurai}, 41

\bibitem[{Lagg} {et~al.}(2010)]{2010ApJ...723L.164L}
{Lagg}, A., {Solanki}, S.~K., {Riethm{\"u}ller}, T.~L., {et~al.} 2010, \apjl,
  723, L164

\bibitem[{Lamb} {et~al.}(2008)]{2008ApJ...674..520L}
{Lamb}, D.~A., {DeForest}, C.~E., {Hagenaar}, H.~J., {Parnell}, C.~E., \&
  {Welsch}, B.~T. 2008, \apj, 674, 520

\bibitem[{Lin} {et~al.}(2025)]{2025AstTI...2..148L}
{Lin}, J., {Feng}, J., {Ge}, Z., {et~al.} 2025, Astronomical Techniques and
  Instruments, 2, 148

\bibitem[{Lites}(2011)]{2011ApJ...737...52L}
{Lites}, B.~W. 2011, \apj, 737, 52

\bibitem[{Lites} {et~al.}(2014)]{2014PASJ...66S...4L}
{Lites}, B.~W., {Centeno}, R., \& {McIntosh}, S.~W. 2014, \pasj, 66, S4

\bibitem[{Lites} {et~al.}(1996)]{1996ApJ...460.1019L}
{Lites}, B.~W., {Leka}, K.~D., {Skumanich}, A., {Martinez Pillet}, V., \&
  {Shimizu}, T. 1996, \apj, 460, 1019

\bibitem[{Lites} {et~al.}(2008)]{2008ApJ...672.1237L}
{Lites}, B.~W., {Kubo}, M., {Socas-Navarro}, H., {et~al.} 2008, \apj, 672, 1237

\bibitem[{Liu} {et~al.}(2025)]{2025ApJ...979..139L}
{Liu}, J., {Sun}, X., {Schuck}, P.~W., \& {Jaeggli}, S.~A. 2025, \apj, 979, 139

\bibitem[{Liu} {et~al.}(2019)]{2019AGUFMSM13B..09L}
{Liu}, J., {Wang}, W., {Thomas}, E.~G., {et~al.} 2019, in AGU Fall Meeting
  Abstracts, Vol. 2019, AGU Fall Meeting Abstracts, SM13B

\bibitem[{Livi} {et~al.}(1985)]{1985MPARp.212..178L}
{Livi}, S.~H.~B., {Wang}, J., \& {Martin}, S.~F. 1985, Max Planck Institut fur
  Astrophysik Report, 212, 178

\bibitem[{Livingston}(1962)]{1962AEEP...16..431L}
{Livingston}, W.~C. 1962, Advances in Electronics and Electron Physics, 16, 431

\bibitem[{Martin}(1988)]{1988cait.rept.....M}
{Martin}, S.~F. 1988, {Studies of flares and disappearing magnetic flux}, Final
  Report, 1 Oct. 1986 - 31 Mar. 1988 California Inst. of Tech., Pasadena. Solar
  Astronomy Group.

\bibitem[{Mart{\'\i}nez Gonz{\'a}lez}
  {et~al.}(2008{\natexlab{a}})]{2008A&A...479..229M}
{Mart{\'\i}nez Gonz{\'a}lez}, M.~J., {Asensio Ramos}, A., {L{\'o}pez Ariste},
  A., \& {Manso Sainz}, R. 2008{\natexlab{a}}, \aap, 479, 229

\bibitem[{Mart{\'\i}nez Gonz{\'a}lez} \& {Bellot
  Rubio}(2009)]{2009ApJ...700.1391M}
{Mart{\'\i}nez Gonz{\'a}lez}, M.~J., \& {Bellot Rubio}, L.~R. 2009, \apj, 700,
  1391

\bibitem[{Mart{\'\i}nez Gonz{\'a}lez} {et~al.}(2012)]{2012ApJ...758L..40M}
{Mart{\'\i}nez Gonz{\'a}lez}, M.~J., {Bellot Rubio}, L.~R., {Solanki}, S.~K.,
  {et~al.} 2012, \apjl, 758, L40

\bibitem[{Mart{\'\i}nez Gonz{\'a}lez}
  {et~al.}(2008{\natexlab{b}})]{2008A&A...477..953M}
{Mart{\'\i}nez Gonz{\'a}lez}, M.~J., {Collados}, M., {Ruiz Cobo}, B., \&
  {Beck}, C. 2008{\natexlab{b}}, \aap, 477, 953

\bibitem[{Mart{\'\i}nez Gonz{\'a}lez} {et~al.}(2010)]{2010ApJ...714L..94M}
{Mart{\'\i}nez Gonz{\'a}lez}, M.~J., {Manso Sainz}, R., {Asensio Ramos}, A., \&
  {Bellot Rubio}, L.~R. 2010, \apjl, 714, L94

\bibitem[{Mart{\'\i}nez Gonz{\'a}lez} {et~al.}(2016)]{2016A&A...596A...5M}
{Mart{\'\i}nez Gonz{\'a}lez}, M.~J., {Pastor Yabar}, A., {Lagg}, A., {et~al.}
  2016, \aap, 596, A5

\bibitem[{Mart{\'\i}nez Pillet}(2013)]{2013SSRv..178..141M}
{Mart{\'\i}nez Pillet}, V. 2013, \ssr, 178, 141

\bibitem[{Martinez Pillet}(2017)]{2017nsf....1718947M}
{Martinez Pillet}, V. 2017, {High Resolution Solar Physics: Past, Present,
  Future}, NSF Award Number 1718947. Directorate for Mathematical and Physical
  Sciences, Division Of Astronomical Sciences. 2017.

\bibitem[{McIntosh} {et~al.}(2011)]{2011Natur.475..477M}
{McIntosh}, S.~W., {de Pontieu}, B., {Carlsson}, M., {et~al.} 2011, \nat, 475,
  477

\bibitem[{Meneguzzi} {et~al.}(1981)]{1981PhRvL..47.1060M}
{Meneguzzi}, M., {Frisch}, U., \& {Pouquet}, A. 1981, \prl, 47, 1060

\bibitem[{Meunier}(2003)]{2003A&A...405.1107M}
{Meunier}, N. 2003, \aap, 405, 1107

\bibitem[{Moll} {et~al.}(2011)]{2011ApJ...736...36M}
{Moll}, R., {Pietarila Graham}, J., {Pratt}, J., {et~al.} 2011, \apj, 736, 36

\bibitem[{Muller} \& {Roudier}(1984)]{1984SoPh...94...33M}
{Muller}, R., \& {Roudier}, T. 1984, \solphys, 94, 33

\bibitem[{Nelson} {et~al.}(2017)]{2017ApJ...845...16N}
{Nelson}, C.~J., {Freij}, N., {Reid}, A., {et~al.} 2017, \apj, 845, 16

\bibitem[{Nelson} {et~al.}(2013)]{2013ApJ...779..125N}
{Nelson}, C.~J., {Shelyag}, S., {Mathioudakis}, M., {et~al.} 2013, \apj, 779,
  125

\bibitem[{Ni} {et~al.}(2022)]{2022A&A...665A.116N}
{Ni}, L., {Cheng}, G., \& {Lin}, J. 2022, \aap, 665, A116

\bibitem[{Ni} {et~al.}(2020)]{2020RSPSA.47690867N}
{Ni}, L., {Ji}, H., {Murphy}, N.~A., \& {Jara-Almonte}, J. 2020, Proceedings of
  the Royal Society of London Series A, 476, 20190867

\bibitem[{Nordlund} {et~al.}(1992)]{1992ApJ...392..647N}
{Nordlund}, A., {Brandenburg}, A., {Jennings}, R.~L., {et~al.} 1992, \apj, 392,
  647

\bibitem[{Orozco Su{\'a}rez} \& {Bellot Rubio}(2012)]{2012ApJ...751....2O}
{Orozco Su{\'a}rez}, D., \& {Bellot Rubio}, L.~R. 2012, \apj, 751, 2

\bibitem[{Orozco Su{\'a}rez} {et~al.}(2012)]{2012ApJ...758L..38O}
{Orozco Su{\'a}rez}, D., {Katsukawa}, Y., \& {Bellot Rubio}, L.~R. 2012, \apjl,
  758, L38

\bibitem[{Orozco Su{\'a}rez} {et~al.}(2007)]{2007ApJ...670L..61O}
{Orozco Su{\'a}rez}, D., {Bellot Rubio}, L.~R., {del Toro Iniesta}, J.~C.,
  {et~al.} 2007, \apjl, 670, L61

\bibitem[{Parker}(1955)]{1955ApJ...121..491P}
{Parker}, E.~N. 1955, \apj, 121, 491

\bibitem[{Petrovay} \& {Szakaly}(1993)]{1993A&A...274..543P}
{Petrovay}, K., \& {Szakaly}, G. 1993, \aap, 274, 543

\bibitem[{Pietarila Graham} {et~al.}(2010)]{2010ApJ...714.1606P}
{Pietarila Graham}, J., {Cameron}, R., \& {Sch{\"u}ssler}, M. 2010, \apj, 714,
  1606

\bibitem[{Pietarila Graham} {et~al.}(2009)]{2009ApJ...693.1728P}
{Pietarila Graham}, J., {Danilovic}, S., \& {Sch{\"u}ssler}, M. 2009, \apj,
  693, 1728

\bibitem[{Pipin} \& {Kosovichev}(2014)]{2014ApJ...785...49P}
{Pipin}, V.~V., \& {Kosovichev}, A.~G. 2014, \apj, 785, 49

\bibitem[{Ponty} {et~al.}(2004)]{2004PhRvL..92n4503P}
{Ponty}, Y., {Politano}, H., \& {Pinton}, J.-F. 2004, \prl, 92, 144503

\bibitem[{Priest}(2014)]{2014masu.book.....P}
{Priest}, E. 2014, {Magnetohydrodynamics of the Sun}

\bibitem[{Rempel}(2014)]{2014ApJ...789..132R}
{Rempel}, M. 2014, \apj, 789, 132

\bibitem[{Rempel}(2018)]{2018ApJ...859..161R}
{Rempel}, M. 2018, \apj, 859, 161

\bibitem[{Rempel}(2020)]{2020ApJ...894..140R}
{Rempel}, M. 2020, \apj, 894, 140

\bibitem[{Rempel} {et~al.}(2023)]{2023SSRv..219...36R}
{Rempel}, M., {Bhatia}, T., {Bellot Rubio}, L., \& {Korpi-Lagg}, M.~J. 2023,
  \ssr, 219, 36

\bibitem[{Rempel} {et~al.}(2009)]{2009ApJ...691..640R}
{Rempel}, M., {Sch{\"u}ssler}, M., \& {Kn{\"o}lker}, M. 2009, \apj, 691, 640

\bibitem[{Requerey} {et~al.}(2014)]{2014ApJ...789....6R}
{Requerey}, I.~S., {Del Toro Iniesta}, J.~C., {Bellot Rubio}, L.~R., {et~al.}
  2014, \apj, 789, 6

\bibitem[{Requerey} {et~al.}(2015)]{2015ApJ...810...79R}
{Requerey}, I.~S., {Del Toro Iniesta}, J.~C., {Bellot Rubio}, L.~R., {et~al.}
  2015, \apj, 810, 79

\bibitem[{Riethm{\"u}ller} {et~al.}(2010)]{2010ApJ...723L.169R}
{Riethm{\"u}ller}, T.~L., {Solanki}, S.~K., {Mart{\'\i}nez Pillet}, V.,
  {et~al.} 2010, \apjl, 723, L169

\bibitem[{Rincon}(2019)]{2019JPlPh..85d2001R}
{Rincon}, F. 2019, Journal of Plasma Physics, 85, 205850401

\bibitem[{Rincon} {et~al.}(2025)]{2025A&A...696A.143R}
{Rincon}, F., {Barr{\`e}re}, P., \& {Roudier}, T. 2025, \aap, 696, A143

\bibitem[{Rodr{\'\i}guez-G{\'o}mez} {et~al.}(2024)]{2024ApJ...964...27R}
{Rodr{\'\i}guez-G{\'o}mez}, J.~M., {Kuckein}, C., {Gonz{\'a}lez Manrique},
  S.~J., {et~al.} 2024, \apj, 964, 27

\bibitem[{Samanta} {et~al.}(2019)]{2019Sci...366..890S}
{Samanta}, T., {Tian}, H., {Yurchyshyn}, V., {et~al.} 2019, Science, 366, 890

\bibitem[{S{\'a}nchez Almeida} {et~al.}(2010)]{2010ApJ...715L..26S}
{S{\'a}nchez Almeida}, J., {Bonet}, J.~A., {Viticchi{\'e}}, B., \& {Del Moro},
  D. 2010, \apjl, 715, L26

\bibitem[{S{\'a}nchez Almeida} {et~al.}(2003)]{2003ApJ...597L.177S}
{S{\'a}nchez Almeida}, J., {Dom{\'\i}nguez Cerde{\~n}a}, I., \& {Kneer}, F.
  2003, \apjl, 597, L177

\bibitem[{Sattarov} {et~al.}(2002)]{2002ApJ...564.1042S}
{Sattarov}, I., {Pevtsov}, A.~A., {Hojaev}, A.~S., \& {Sherdonov}, C.~T. 2002,
  \apj, 564, 1042

\bibitem[{Scharmer} {et~al.}(2003)]{2003SPIE.4853..341S}
{Scharmer}, G.~B., {Bjelksjo}, K., {Korhonen}, T.~K., {Lindberg}, B., \&
  {Petterson}, B. 2003, in Society of Photo-Optical Instrumentation Engineers
  (SPIE) Conference Series, Vol. 4853, Innovative Telescopes and
  Instrumentation for Solar Astrophysics, ed. S.~L. {Keil} \& S.~V. {Avakyan},
  341

\bibitem[{Schekochihin} {et~al.}(2004{\natexlab{a}})]{2004Ap&SS.292..141S}
{Schekochihin}, A.~A., {Cowley}, S.~C., {Taylor}, S.~F., {Maron}, J.~L., \&
  {McWilliams}, J.~C. 2004{\natexlab{a}}, \apss, 292, 141

\bibitem[{Schekochihin} {et~al.}(2004{\natexlab{b}})]{2004ApJ...612..276S}
{Schekochihin}, A.~A., {Cowley}, S.~C., {Taylor}, S.~F., {Maron}, J.~L., \&
  {McWilliams}, J.~C. 2004{\natexlab{b}}, \apj, 612, 276

\bibitem[{Schekochihin} {et~al.}(2005)]{2005ApJ...625L.115S}
{Schekochihin}, A.~A., {Haugen}, N.~E.~L., {Brandenburg}, A., {et~al.} 2005,
  \apjl, 625, L115

\bibitem[{Schekochihin} {et~al.}(2007)]{2007NJPh....9..300S}
{Schekochihin}, A.~A., {Iskakov}, A.~B., {Cowley}, S.~C., {et~al.} 2007, New
  Journal of Physics, 9, 300

\bibitem[{Schmidt}(2015)]{2015LRCA....1....2S}
{Schmidt}, W. 2015, Living Reviews in Computational Astrophysics, 1, 2

\bibitem[{Schrijver} \& {Harvey}(1994)]{1994SoPh..150....1S}
{Schrijver}, C.~J., \& {Harvey}, K.~L. 1994, \solphys, 150, 1

\bibitem[{Schrijver} {et~al.}(1998)]{1998Natur.394..152S}
{Schrijver}, C.~J., {Title}, A.~M., {Harvey}, K.~L., {et~al.} 1998, \nat, 394,
  152

\bibitem[{Schumacher} \& {Sreenivasan}(2020)]{2020RvMP...92d1001S}
{Schumacher}, J., \& {Sreenivasan}, K.~R. 2020, Reviews of Modern Physics, 92,
  041001

\bibitem[{Sch{\"u}ssler}(2013)]{2013IAUS..294...95S}
{Sch{\"u}ssler}, M. 2013, in IAU Symposium, Vol. 294, Solar and Astrophysical
  Dynamos and Magnetic Activity, ed. A.~G. {Kosovichev}, E.~{de Gouveia Dal
  Pino}, \& Y.~{Yan}, 95

\bibitem[{Sch{\"u}ssler} \& {V{\"o}gler}(2008)]{2008A&A...481L...5S}
{Sch{\"u}ssler}, M., \& {V{\"o}gler}, A. 2008, \aap, 481, L5

\bibitem[{Shchukina} \& {Trujillo Bueno}(2011)]{2011ApJ...731L..21S}
{Shchukina}, N., \& {Trujillo Bueno}, J. 2011, \apjl, 731, L21

\bibitem[{Shen} {et~al.}(2022)]{2022SoPh..297...20S}
{Shen}, Y., {Zhou}, X., {Duan}, Y., {et~al.} 2022, \solphys, 297, 20

\bibitem[{Smagorinsky}(1963)]{1963MWRv...91...99S}
{Smagorinsky}, J. 1963, Monthly Weather Review, 91, 99

\bibitem[{Smitha} {et~al.}(2017)]{2017ApJS..229...17S}
{Smitha}, H.~N., {Anusha}, L.~S., {Solanki}, S.~K., \& {Riethm{\"u}ller}, T.~L.
  2017, \apjs, 229, 17

\bibitem[{Socas-Navarro} \& {Lites}(2004)]{2004ApJ...616..587S}
{Socas-Navarro}, H., \& {Lites}, B.~W. 2004, \apj, 616, 587

\bibitem[{Solanki} {et~al.}(2010)]{2010ApJ...723L.127S}
{Solanki}, S.~K., {Barthol}, P., {Danilovic}, S., {et~al.} 2010, \apjl, 723,
  L127

\bibitem[{Spitzer}(1962)]{1962pfig.book.....S}
{Spitzer}, L. 1962, {Physics of Fully Ionized Gases}

\bibitem[{Stein} \& {Nordlund}(1998)]{1998ApJ...499..914S}
{Stein}, R.~F., \& {Nordlund}, {\r{A}}. 1998, \apj, 499, 914

\bibitem[{Steiner} \& {Rezaei}(2012)]{2012ASPC..456....3S}
{Steiner}, O., \& {Rezaei}, R. 2012, in Astronomical Society of the Pacific
  Conference Series, Vol. 456, Fifth Hinode Science Meeting, ed. L.~{Golub},
  I.~{De Moortel}, \& T.~{Shimizu}, 3

\bibitem[{Steiner} {et~al.}(2008)]{2008ApJ...680L..85S}
{Steiner}, O., {Rezaei}, R., {Schaffenberger}, W., \& {Wedemeyer-B{\"o}hm}, S.
  2008, \apjl, 680, L85

\bibitem[{Stenflo}(1973)]{1973SoPh...32...41S}
{Stenflo}, J.~O. 1973, \solphys, 32, 41

\bibitem[{Stenflo}(2012)]{2012A&A...541A..17S}
{Stenflo}, J.~O. 2012, \aap, 541, A17

\bibitem[{Stenflo}(2013)]{2013A&A...555A.132S}
{Stenflo}, J.~O. 2013, \aap, 555, A132

\bibitem[{Stenflo} {et~al.}(2002)]{2002A&A...389..314S}
{Stenflo}, J.~O., {Gandorfer}, A., {Holzreuter}, R., {et~al.} 2002, \aap, 389,
  314

\bibitem[{Tarbell} {et~al.}(1979)]{1979ApJ...229..387T}
{Tarbell}, T.~D., {Title}, A.~M., \& {Schoolman}, S.~A. 1979, \apj, 229, 387

\bibitem[{Thornton} \& {Parnell}(2011)]{2011SoPh..269...13T}
{Thornton}, L.~M., \& {Parnell}, C.~E. 2011, \solphys, 269, 13

\bibitem[{Tian} {et~al.}(2014)]{2014ApJ...797L..14T}
{Tian}, H., {Li}, G., {Reeves}, K.~K., {et~al.} 2014, \apjl, 797, L14

\bibitem[{Tobias} {et~al.}(2011)]{2011arXiv1103.3138T}
{Tobias}, S.~M., {Cattaneo}, F., \& {Boldyrev}, S. 2011, arXiv e-prints,
  arXiv:1103.3138

\bibitem[{Toriumi} \& {Yokoyama}(2011)]{2011ApJ...735..126T}
{Toriumi}, S., \& {Yokoyama}, T. 2011, \apj, 735, 126

\bibitem[{Trujillo Bueno} {et~al.}(2004)]{2004Natur.430..326T}
{Trujillo Bueno}, J., {Shchukina}, N., \& {Asensio Ramos}, A. 2004, \nat, 430,
  326

\bibitem[{Tu} {et~al.}(2005)]{2005Sci...308..519T}
{Tu}, C.-Y., {Zhou}, C., {Marsch}, E., {et~al.} 2005, Science, 308, 519

\bibitem[{Utz} {et~al.}(2014)]{2014ApJ...796...79U}
{Utz}, D., {del Toro Iniesta}, J.~C., {Bellot Rubio}, L.~R., {et~al.} 2014,
  \apj, 796, 79

\bibitem[{Vargas Dom{\'\i}nguez} {et~al.}(2014)]{2014ApJ...794..140V}
{Vargas Dom{\'\i}nguez}, S., {Kosovichev}, A., \& {Yurchyshyn}, V. 2014, \apj,
  794, 140

\bibitem[{Vitense}(1953)]{1953ZA.....32..135V}
{Vitense}, E. 1953, \zap, 32, 135

\bibitem[{V{\"o}gler} \& {Sch{\"u}ssler}(2007)]{2007A&A...465L..43V}
{V{\"o}gler}, A., \& {Sch{\"u}ssler}, M. 2007, \aap, 465, L43

\bibitem[{Wang} {et~al.}(1995)]{1995SoPh..160..277W}
{Wang}, J., {Wang}, H., {Tang}, F., {Lee}, J.~W., \& {Zirin}, H. 1995,
  \solphys, 160, 277

\bibitem[{Wang} \& {Zhang}(1999)]{1999ASPC..184..222W}
{Wang}, J., \& {Zhang}, J. 1999, in Astronomical Society of the Pacific
  Conference Series, Vol. 184, Third Advances in Solar Physics Euroconference:
  Magnetic Fields and Oscillations, ed. B.~{Schmieder}, A.~{Hofmann}, \&
  J.~{Staude}, 222

\bibitem[{Warnecke} {et~al.}(2023)]{2023NatAs...7..662W}
{Warnecke}, J., {Korpi-Lagg}, M.~J., {Gent}, F.~A., \& {Rheinhardt}, M. 2023,
  Nature Astronomy, 7, 662

\bibitem[{Zeeman}(1897)]{1897ApJ.....5..332Z}
{Zeeman}, P. 1897, \apj, 5, 332

\bibitem[{Zhang} \& {Zhang}(2000)]{2000SoPh..196..269Z}
{Zhang}, H., \& {Zhang}, M. 2000, \solphys, 196, 269

\bibitem[{Zhang} {et~al.}(2009)]{2009RAA.....9..921Z}
{Zhang}, J., {Yang}, S.-H., \& {Jin}, C.-L. 2009, Research in Astronomy and
  Astrophysics, 9, 921

\bibitem[{Zhang} \& {Zhang}(1999)]{1999A&A...352..317Z}
{Zhang}, M., \& {Zhang}, H.~Q. 1999, \aap, 352, 317

\bibitem[{Zhou} {et~al.}(2013)]{2013SoPh..283..273Z}
{Zhou}, G., {Wang}, J., \& {Jin}, C. 2013, \solphys, 283, 273

\end{thebibliography}

\end{document}